\documentclass[10pt,journal,compsoc]{IEEEtran}

\usepackage[ruled, vlined, linesnumbered]{algorithm2e}
\usepackage{array}
\usepackage{multirow}

\usepackage{cite}
\usepackage{mathtools}
\usepackage{amsmath}
\usepackage{amssymb}
\usepackage{adjustbox}
\usepackage[dvipsnames]{xcolor} 

\allowdisplaybreaks

\usepackage[hidelinks]{hyperref}

\ifCLASSOPTIONcompsoc
    \usepackage[caption=false, font=normalsize, labelfont=sf, textfont=sf]{subfig}
\else
\usepackage[caption=false, font=footnotesize]{subfig}
\fi

\usepackage{footnote}
\usepackage[left=0.75in,right=0.75in,top=1in,bottom=1in]{geometry}
\usepackage{enumitem}
\usepackage{flushend}
\usepackage{courier}
\usepackage{algpseudocode}

\usepackage{cleveref}
\crefformat{section}{\S#2#1#3}
\crefformat{subsection}{\S#2#1#3}
\crefformat{subsubsection}{\S#2#1#3}

\usepackage[nopostdot,nogroupskip,nonumberlist]{glossaries}

\newcommand{\toolname}{{Sensifi}}

\newacronym{IoT}{IoT}{Internet of Things}

\newacronym{SCU}{SCU}{Santa Clara University}
\newacronym{UTM}{UTM}{University Technology Malaysia}
\newacronym{SES}{SES}{Summer Engineering Seminar}

\newacronym{COTS}{COTS}{Commercial Off-The-Shelf}

\newacronym{MQTT}{MQTT}{Message Queue Telemetry Transport}

\newacronym{API}{API}{Application Programming Interface}

\newacronym{CSMA}{CSMA}{Carrier-Sense Multiple Access}
\newacronym{TDMA}{TDMA}{Time-Division Multiple Access}

\newacronym{DMA}{DMA}{Direct Memory Access}

\newacronym{MTU}{MTU}{Maximum Transmission Unit}

\newacronym{sps}{sps}{samples per second}

\newacronym{TST}{TST}{Time-stamp Table}
\newacronym{TIT}{TIT}{Time-stamp Indexing Table}
\newacronym{GSFC}{GSFC}{Goddard Space Flight Center}
\newacronym{JPL}{JPL}{Jet Propulsion Laboratory}

\newacronym{EDF}{EDF}{Earliest Deadline First}

\newacronym{LDPC}{LDPC}{Low-Density Parity-Check}
\newacronym{STBC}{STBC}{Space-Time Block Coding }
\newacronym{GI}{GI}{Guard Interval}

\newacronym{WAIC}{WAIC}{Wireless Avionics Intra-Communications}

\newacronym{I2A}{I2A}{Idle to Alert}
\newacronym{A2S}{A2S}{Alert to Sampling}

\newacronym{TSF}{TSF}{Time Synchronization Function}

\newacronym{TSM}{TSM}{Time Synchronization Method}

\newglossaryentry{RTOS}
{
  name={RTOS},
  description={Real-Time Operating System},
  first={\glsentrydesc{RTOS} (\glsentrytext{RTOS})},
  plural={RTOSes},
  firstplural={\glsentrydesc{RTOS}s (\glsentryplural{RTOS})}
}

\newglossaryentry{SoC}
{
  name={SoC},
  description={System on a Chip},
  first={\glsentrydesc{SoC} (\glsentrytext{SoC})},
  plural={SoCs},
  firstplural={\glsentrydesc{SoC}s (\glsentryplural{SoC})}
}

\newglossaryentry{WSN}
{
  name={WSN},
  description={Wireless Sensor Network},
  first={\glsentrydesc{WSN} (\glsentrytext{WSN})},
  plural={WSNs},
  firstplural={\glsentrydesc{WSN}s (\glsentryplural{WSN})}
}

\newglossaryentry{ADC}
{
  name={ADC},
  description={Analog to Digital Converter},
  first={\glsentrydesc{ADC} (\glsentrytext{ADC})},
  plural={ADCs},
  firstplural={\glsentrydesc{ADC}s (\glsentryplural{ADC})}
}

\newglossaryentry{DAC}
{
  name={DAC},
  description={Digital to Analog Converter},
  first={\glsentrydesc{DAC} (\glsentrytext{DAC})},
  plural={DACs},
  firstplural={\glsentrydesc{DAC}s (\glsentryplural{DAC})}
}

\newglossaryentry{DAQ}
{
  name={DAQ},
  description={Data Acquisition},
  first={\glsentrydesc{DAQ} (\glsentrytext{DAQ})},
  plural={DAQs},
  firstplural={\glsentrydesc{DAQ}s (\glsentryplural{DAQ})}
}

\newacronym{SUT}{SUT}{System Under Test}

\newacronym{LDO}{LDO}{low-dropout}

\newglossaryentry{qdisc}
{
  name={qdisc},
  description={queuing discipline},
  first={\glsentrydesc{qdisc} (\glsentrytext{qdisc})},
  plural={qdiscs},
  firstplural={\glsentrydesc{qdisc}s (\glsentryplural{qdisc})}
}

\newacronym{SAR}{SAR}{Successive Approximation Register}

\newacronym{SPI}{SPI}{Serial Peripheral Interface}
\newacronym{NTP}{NTP}{Network Time Protocol}

\newacronym{IENC}{IENC}{Interval Encoding}
\newacronym{OENC}{OENC}{Outlier Encoding}
\newacronym{D-OENC}{D-OENC}{Differential Outlier Encoding}

\newacronym{MEMS}{MEMS}{Micro-Electro-Mechanical Systems}

\newacronym{MAE}{MAE}{Mean Absolute Error}

\newacronym{RFR}{RFR}{Random Forest Regressor}
\newacronym{GBR}{GBR}{Gradient Boosting Regressor}
\newacronym{ABR}{ABR}{Adaptive Boosting Regressor}
\newacronym{ETR}{ETR}{Extra Trees Regressor}
\newacronym{LSTM}{LSTM}{Long Short-Term Memory}
\newacronym{HBR}{HBR}{Histogram-Based Gradient Boosting Regressor}

\newacronym{H2H}{H2H}{Humand to Human}
\newacronym{M2H}{M2H}{Machine to Human}
\newacronym{M2M}{M2M}{Machine to Machine}

\newacronym{MAC}{MAC}{Medium Access Control}
\newacronym{CW}{CW}{Contention Window}
\newacronym{MLME}{MLME}{MAC layer management entity }

\newacronym{BAS}{BAS}{Building Automation System}
\newacronym{PA}{PA}{Process Automation}
\newacronym{FA}{FA}{Factory Automation}
\newacronym{PSA}{PSA}{Power Systems Automation}
\newacronym{PEC}{PEC}{Power Electronics Control}

\newacronym{RAT}{RAT}{Radio Access Technology}

\newacronym{CAGR}{CAGR}{Compound Annual Growth Rate}

\newacronym{RSSI}{RSSI}{Received Signal Strength Indicator}
\newacronym{SNR}{SNR}{Singal-to-Noise Ratio}

\newacronym{MCS}{MCS}{Modulation Coding Scheme}
\newacronym{MIMO}{MIMO}{Multiple-Input Multiple-Output}
\newacronym{PER}{PER}{Packet Error Rate}
\newacronym{MPDU}{MPDU}{MAC protocol data unit}
\newacronym{A-MPDU}{A-MPDU}{Aggregated-MAC protocol data unit}

\newacronym{DUT}{DUT}{Device Under Test}
\newacronym{BLE}{BLE}{Bluetooth Low Energy}

\newacronym{NFC}{NFC}{Near Field Communication}

\newacronym{WEP}{WEP}{Wired Equivalent Privacy }
\newacronym{WPA}{WPA}{Wireless Protected Access}
\newacronym{PMK}{PMK}{Pairwise Master Key}
\newacronym{PTK}{PTK}{Pairwise Transition Keys}
\newacronym{GTK}{GTK}{Group Temporal Key}
\newacronym{EAP}{EAP}{Extensible Authentication Protocol}
\newacronym{RRM}{RRM}{Radio Resource Management}
\newacronym{BSSTM}{BSSTM}{BSS Transition Management}
\newacronym{BSS}{BSS}{Basic Service Set}
\newacronym{SSID}{SSID}{Service Set Identifier}
\newacronym{BSSID}{BSSID}{BSS Identifier}
\newacronym{CSA}{CSA}{Channel Switching Announcement}
\newacronym{DCF}{DCF}{Distributed Coordination Function}
\newacronym{EDCF}{EDCF}{Enhanced Distributed Coordination Function}
\newacronym{AIFS}{AIFS}{Arbitration Inter-Frame Space}

\newacronym{QoS}{QoS}{Quality of Service}
\newacronym{ToS}{ToS}{Type of Service}

\newacronym{AC}{AC}{Access Category}
\newacronym{PSM}{PSM}{Power Save Mode}
\newacronym{PSP}{PSP}{Power Save Poll}
\newacronym{PSP-N}{PSP-N}{Power Save Poll with Null}
\newacronym{CAM}{CAM}{Continuously Active Mode}
\newacronym{AID}{AID}{Association ID}
\newacronym{APSM}{APSM}{Adaptive PSM}
\newacronym{APSD}{APSD}{Automatic Power Save Delivery}
\newacronym{U-APSD}{U-APSD}{Unscheduled-Automatic Power Save Delivery}
\newacronym{S-APSD}{S-APSD}{Scheduled-Automatic Power Save Delivery}
\newacronym{EOSP}{EOSP}{End of Service Period}
\newacronym{SP}{SP}{Service Period}
\newacronym{TXOP}{TXOP}{Transmit Opportunity}
\newacronym{BI}{BI}{Beacon Interval}

\newacronym{DBS}{DBS}{Disassociation-Based Steering}

\newacronym{SNMP}{SNMP}{Simple Network Management Protocol}
\newacronym{MIB}{MIB}{Management Information Base}

\newacronym{ODL}{ODL}{OpenDayLight} 
\newacronym{OVS}{OVS}{Open vSwitch}
\newacronym{OVSDB}{OVSDB}{Open vSwitch Database}
\newacronym{SDN}{SDN}{Software Defined Networking}

\newacronym{VM}{VM}{Virtual Machine}

\newacronym{NF}{NF}{Network Function}
\newacronym{VNFM}{VNFM}{VNF Manager}
\newacronym{VNF}{VNF}{Virtual Network Function}
\newacronym{NNF}{NNF}{Native Network Function}
\newacronym{NFVO}{NFVO}{NFV Orchestrator}
\newacronym{VIM}{VIM}{Virtualized Infrastructure Manager}

\newacronym{FQ-CoDel}{FQ-CoDel}{Flow Queuing with Controlled Delay}
\newacronym{HTB}{HTB}{Hierarchical Token Bucket}
\newacronym{FQ}{FQ}{Fair Queuing}
\newacronym{SFQ}{SFQ}{Stochastic Fair Queuing}
\newacronym{BQL}{BQL}{Buffer Queue Limits}

\newacronym{PCP}{PCP}{Priority Code Point}

\newacronym{NAT}{NAT}{Network Address Translation}

\newacronym{NUMA}{NUMA}{Non-Uniform Memory Access}

\newacronym{CPE}{CPE}{Customer Premise Equipment}

\newacronym{DSCP}{DSCP}{Differentiated Services Code Point}

\newacronym{vSG}{vSG}{Virtual Subscriber Gateway}

\newacronym{RGW}{RGW}{Residential Gateway}
\newacronym{vRGW}{vRGW}{Virtual RGW}

\newacronym{VF}{VF}{Virtual Function}
\newacronym{PF}{PF}{Physical Function}

\newacronym{ARP}{ARP}{Address Resolution Protocol}

\newacronym{TIM}{TIM}{Traffic Indication Message}
\newacronym{DTIM}{DTIM}{Delivery Traffic Indication Message}
\newacronym{AP}{AP}{Access Point}
\newacronym{STA}{STA}{Station}
\newacronym{TWT}{TWT}{Target Wake-up Time}
\newacronym{OPS}{OPS}{Opportunistic Power Save}
\newacronym{PCF}{PCF}{Point Coordination Function}
\newacronym{PIFS}{PIFS}{PCP Inter Frame Space}
\newacronym{DIFS}{DIFS}{Distributed Inter Frame Space}
\newacronym{BSR}{BSR}{Buffer Status Report}
\newacronym{OCW}{OCW}{OFDM Contention Window}
\newacronym{OFDMA}{OFDMA}{Orthogonal Frequency-Division Multiple Access}
\newacronym{OBO}{OBO}{OFDMA Back-off}
\newacronym{RU}{RU}{Resource Unit}

\newacronym{QTP}{QTP}{Quite Time Period}
\newacronym{AU}{AU}{Airtime Utilization}

\newacronym{LVAP}{LVAP}{Light Virtual AP}
\newacronym{EDCA}{EDCA}{Enhanced Distributed Channel Access}

\newacronym{HCCA}{HCCA}{Hybrid Controlled Channel Access}
\newacronym{AR-ARX}{AR-ARX}{Autoregressive-Autoregressive Exogenous}
\newacronym{WOL}{WOL}{Wake-on-Lan}
\newacronym{DMM}{DMM}{Digital Multimeter}

\newacronym{DC}{DC}{Datacenter}
\newacronym{CO}{CO}{Central Office}

\newacronym{CORD}{CORD}{Central Office Re-architected as a Datacenter}
\newacronym{MEC}{MEC}{Mobile Edge Computing}
\newacronym{ASP}{ASP}{Application Service Provider}
\newacronym{ES}{ES}{Edge Service}
\newacronym{NSP}{NSP}{Network Service Provider}
\newacronym{PON}{PON}{Passive Optical Network}
\newacronym{MSMA}{MSMA}{Multiple Subcarrier Multiple Access}
\newacronym{NV}{NV}{Network Virtualization}
\newacronym{NFV}{NFV}{Network Function Virtualization}
\newacronym{NFVI}{NFV}{NFV Infrastructure}
\newacronym{MANO}{MANO}{Management and Orchestration}

\newacronym{NIC}{NIC}{Network Interface Card}
\newacronym{vNIC}{vNIC}{virtual NIC}
\newacronym{pNIC}{pNIC}{physical NIC}

\newacronym{TCAM}{TCAM}{Tenary Content Addressable Memory}
\newacronym{RSS}{RSS}{Receive Side Scaling}

\newacronym{DPDK}{DPDK}{Data Plane Development Kit}
\newacronym{UIO}{UIO}{User-space I/O}

\newacronym{EMC}{EMC}{Exact Match Cache}

\newacronym{SR-IOV}{SR-IOV}{Single Root-Input/Output Scaling}
\newacronym{PMD}{PMD}{Poll Mode Driver}

\newacronym{TSS}{TSS}{Tuple Space Search}
\newacronym{dpcls}{dpcls}{data path classifier}
\newacronym{GVL}{GVL}{General Vibration Laboratory}
\newacronym{FDK}{FDK}{Fog Development Kit}
\newacronym{RAA}{RAA}{Resource Allocation Algorithm}

\newacronym{TSN}{TSN}{Time Sensitive Networking}
\newacronym{AFDX}{AFDX}{avionics full-duplex switched Ethernet}
\newacronym{CAN}{CAN}{Controller Area Network}

\newacronym{OCI}{OCI}{Open Carrier Interface }

\newacronym{CBR}{CBR}{constant bit rate}

\newacronym{EWU}{EWU}{Early Wake-UP}
\newacronym{LWU}{LWU}{Late Wake-UP}
\newacronym{MWU}{MWU}{Mid Wake-UP}

\newacronym{DP}{DP}{Data Path}
\newacronym{FCC}{FCC}{Federal Communications Commission}
\newacronym{MPTCP}{MPTCP}{MultiPath TCP}

\newacronym{HAL}{HAL}{Hardware Abstraction Layer}

\newacronym{VFS}{VFS}{Virtual File System}

\newacronym{NOTP}{NOTP}{Name Of The Paper}
\newacronym{SHM}{SHM}{Structural Health Monitoring}
\newacronym{IBSS}{IBSS}{Independent Basic Service Set}
\newacronym{SATSF}{SATSF}{Self-Adjusting Timing Synchronization Function}
\newacronym{MATSF}{MATSF}{Multihop Adaptive Timing Synchronization Function}
\newacronym{DW}{DW}{Delayed Wake-up}
\newacronym{ROD}{ROD}{Radio-On-Demand}

\ifCLASSINFOpdf
\else
\fi

\begin{document}



%


\title{Sensifi: A Wireless Sensing System for Ultra-High-Rate Applications}


\twocolumn

%
%
%
%


\author{\IEEEauthorblockN{Chia-Chi~Li, Vikram~K.~Ramanna, Daniel~Webber, Cole~Hunter, Tyler~Hack, and Behnam Dezfouli\IEEEauthorrefmark{1}\thanks{\IEEEauthorrefmark{1}Corresponding Author.}}\\
\IEEEauthorblockA{\small Internet of Things Research Lab, Santa Clara University, USA\\
\texttt{\{cli1, vramanna, dwebber, chunter, thack, bdezfouli\}@scu.edu}\\
}
}

\IEEEtitleabstractindextext{
\begin{abstract}
Wireless Sensor Networks (WSNs) are being used in various applications such as structural health monitoring and industrial control.
Since energy efficiency is one of the major design factors, the existing WSNs primarily rely on low-power, \textit{low-rate} wireless technologies such as 802.15.4 and Bluetooth.
In this paper, by proposing \toolname{}, we strive to tackle the challenges of developing \textit{ultra-high-rate} WSNs based on the 802.11 (WiFi) standard.
As an illustrative structural health monitoring application, we consider spacecraft vibration test and identify system design requirements and challenges.
Our main contributions are as follows.
First, we propose packet encoding methods to reduce the overhead of assigning accurate timestamps to samples.
Second, we propose energy efficiency methods to enhance the system's lifetime.
Third, to enhance sampling rate and mitigate sampling rate instability, we reduce the overhead of processing outgoing packets through the network stack.
Fourth, we study and reduce the delay of processing time synchronization packets through the network stack.
Fifth, we propose a low-power node design particularly targeting vibration monitoring.
Sixth, we use our node design to empirically evaluate energy efficiency, sampling rate, and data rate.
We leave large-scale evaluations as future work.


\end{abstract}

}

\maketitle

\pagenumbering{arabic}
\thispagestyle{plain}
\pagestyle{plain}

\IEEEdisplaynontitleabstractindextext

%
\IEEEpeerreviewmaketitle

\section{Introduction}
\label{sec:introduction}

The reductions in the cost and energy consumption of microprocessors, wireless transceivers and sensors have facilitated the development of \glspl{WSN} for \gls{SHM} as well as applications such as \gls{PA}, \gls{FA}, and medical monitoring~\cite{dezfouli2017rewimo,notay2011wireless,huang2015development,girard2013environmental,loutas2012intelligent}.
\glspl{WSN} are being used in \gls{SHM} applications to monitor spacecraft, aerial vehicles, high-rise buildings, dams, highways, and bridges~\cite{wijetunge2010wireless,notay2011wireless,liu2016senetshm,harms2010structural}.
In these applications, sensor nodes collect information such as acceleration, ambient vibration, load, and stress at sampling frequencies usually above $\textrm{100\:Hz}$~\cite{xu2004wireless}.
The replacement of cables with wireless links reduces design, deployment, and maintenance costs.
Also, wireless links offer higher reliability compared to cables that are subject to wear and tear.

In this paper, we use the term \textit{ultra-high-rate} to refer to the systems whose data transmission rate is in the range of few hundreds of Mbps to a few Gbps.
The ultra-high-rate of communication is the result of high-rate, high-resolution sampling.
Applications such as \gls{FA}, \gls{PSA}, and \gls{PEC} are representative of ultra-high-rate scenarios \cite{luvisotto2016ultra}.
A prominent example of \gls{SHM} is spacecraft vibration monitoring.
These tests are performed to analyze a spacecraft's mechanical durability against the launch's powerful vibrational forces.
These tests are conducted with a large number of wired accelerometers (more than 200), simultaneously sampled by expensive, high-performance \gls{DAQ} modules~\cite{fantasia2011vibration}.
The large number of nodes and the high sampling rate (around $\textrm{50\:kHz}$) result in generating a massive amount of data that must be collected from the structure.
This results in a disarray of wired connections that make installing and debugging these sensors an excessively laborious and time-consuming process.
In addition to the setup difficulties, the extra weight applied to the spacecraft by the cables can skew vibrational measurement results~\cite{whatzwrong}.
The cables also introduce electrostatic noise that affects data collection accuracy.

Besides satisfying the high sampling rate required by these applications, the additional system requirements are:
(\romannumeral 1) \textit{liveness}: collecting data from all the nodes while sampling is in progress, 
(\romannumeral 2) \textit{scalability}: live data collection from a large number of nodes (e.g., more than 200 in spacecraft monitoring),
(\romannumeral 3) \textit{energy efficiency}: the nodes' longevity must satisfy the duration requirement of the experiment,
(\romannumeral 4) \textit{accuracy}: accurate timestamps must be assigned to the samples,
and 
(\romannumeral 5) \textit{reliability}: lossless collection of raw samples is required.
%
Unfortunately, the existing systems do not address these requirements.
We highlight design challenges and the shortcomings of existing systems as follows.
\textit{First}, considering the energy limitations of sensor nodes, 802.15.4 and 802.15.1 are the primary wireless technologies used in existing works \cite{dezfouli2015modeling}.
Consider a node collecting $\textrm{12-bit}$ samples at $\textrm{50\:k\gls{sps}}$. 
This node needs to communicate at $\textrm{600\:kbps}$, excluding the overhead of the samples' timestamps; hence, the $\textrm{250\:kbps}$ data rate provided by the 802.15.4 standard is less than the communication demand of \textit{one node}.
Similarly, the wireless technologies used in current industrial networks, such as WirelessHART~\cite{hart2020hart}, WIA-PA~\cite{iec_WIA_PA}, ISA 100~\cite{isa100}, WSAN-FA/IO-Link~\cite{heynicke2017io}, and WISA~\cite{scheible2007unplugged} cannot be used for ultra-high-rate sensing applications.
\textit{Second}, the liveness requirement is not satisfied if samples are stored in the main memory or a memory card and then transferred whenever the user requests for data delivery~\cite{valenti2018low,kim2007health,araujo2011wireless,woo2003taming}.
For example, spacecraft vibration tests require live response monitoring, identifying malfunctioning nodes, and making adjustments to the test configuration parameters.
Besides, frequent placement and removal of nodes in hard-to-access areas are not desirable.
\textit{Third}, compression and in-network processing methods are lossy and cannot be used in scenarios where raw data collection is required to implement various data analysis methods~\cite{liu2016senetshm,werner2006deploying}.
Furthermore, the compression level provided by these methods is not enough to adapt low-rate wireless technologies for ultra-high-rate applications.
Also, the high processing overhead of these compression algorithms causes a considerable interruption in sample collection.
\textit{Fourth}, variations of inter-sample delay make assigning accurate timestamps to samples more challenging, and achieving sampling rate stability is particularly difficult in ultra-high-rate applications.
Running additional tasks on the processor causes interruptions of sample collection and variations of the sampling rate.
Therefore, the effect of running additional tasks, such as packet processing and wireless transmission, must be minimized.
The existing work has identified this challenge; however, they do not offer a comprehensive evaluation or mitigation of these problems~\cite{kim2007health,whelan2009real}.

In this paper, we propose \toolname{}, a system based on 802.11 (WiFi) standard for ultra-high-rate sensing applications.
Towards tackling the underlying challenges of developing this system, \textit{the main contributions of this paper are as follows:}
\textit{First} (\cref{sample_coll_and_enc}), we show that assigning accurate timestamps ($\mathrm{sub-\mu s}$ accuracy) to samples introduces a significant overhead and wastes precious payload bytes and wireless communication bandwidth.
Considering the distribution of inter-sample intervals, we propose three encoding methods to reduce payload overhead. 
These methods achieve various levels of payload efficiency and execution time.
By reducing the amount of data sent per node, we enhance system scalability and reduce the nodes' energy consumption to transmit the collected samples.
\textit{Second} (\cref{section:idle_alert_phase}), we propose methods to reduce the energy consumption of nodes during their inactivity periods.
In particular, since the 802.11 transceiver is the primary energy consumption source, we study two energy-efficiency methods: periodic association with the \gls{AP}, and extended-period beacon reception.
We also propose and empirically evaluate mechanisms to enhance the energy efficiency of these methods further.
Depending on the application scenario, the proposed methods and configurations can be used to achieve the desired energy-delay tradeoff.
\textit{Third} (\cref{section:packet_processing}), we show that preparing and processing outgoing packets by the network stacks introduce a non-negligible delay that must be minimized to increase sampling rate and stability.
Our work reveals that bypassing transport layer processing is essential to minimize the effect of packet processing on sampling rate, and this bypassing also translates into the higher performance of timestamp encoding.
\textit{Fourth} (\cref{time_sync}), we present a thorough study of time synchronization and identify the challenges caused by network stack complexity.
We place the time synchronization function in the 802.11 subsystem, between the packet decoding logic and driver, and confirm that the maximum error is bounded by $\mathrm{0.25\:\mu s}$.
\textit{Fifth} (\cref{hardware_design}), we present a modular, low-power node design. 
This node is capable of collecting 16-bit samples at rates up to $\textrm{500\:ksps}$.
Therefore, it enables us to measure sampling rate variations, study the impact of packet processing on the sampling rate, and evaluate energy consumption.
Compared with the existing works that rely on TinyOS~\cite{liu2016senetshm,whelan2009real,alves2017damage}, Linux~\cite{wei2013rt,seno2016enhancing,tramarin2019real}, or theoretical performance analysis~\cite{islam2018structural,ling2009localized}, our hardware platform supports the two widely-used \glspl{RTOS} targeting resource-constrained devices: FreeRTOS~\cite{FreeRTOS,FreeRTOS_git} and ThreadX~\cite{ThreadX,ThreadX_github}.
Also, we use the network stacks available for these \glspl{RTOS}, namely, LwIP~\cite{LwIP}, NetX~\cite{NetX}, and NetXDuo~\cite{NetXDuo}.
Therefore, a unique aspect of our work is identifying and tackling the underlying challenges of using these tools for developing ultra-high-rate applications.
\textit{Sixth} (\cref{perf_eval}), using our node design, we empirically evaluate various aspects of the system and demonstrate the performance enhancements achieved with the proposed methods.
However, we leave large-scale throughput evaluation of the system as future work.
The methods proposed in this paper can be employed for developing \gls{WSN} systems for applications such as \gls{PA}, \gls{FA}, \gls{BAS}, intra-vehicle communication, and \gls{WAIC}.

\section{Sample Application Scenario}
\label{app_scenario}

Although the proposed system can be used in various ultra-high-rate sensing applications, we overview spacecraft vibration monitoring to identify and clarify the unique challenges of this application and frame our system design.

Spacecraft vibration test monitoring is a \gls{SHM} application that requires live, ultra-high-rate data collection from a large number of nodes (usually more than 200)~\cite{chang2000deep,kolaini2018spacecraft}.
The main objective of this test is to excite components, subsystems, and nonstructural hardware that respond to the launch environments.
The spacecraft is attached to a large shake table, and then, accelerometers are mounted. 
This process is referred to as the \textit{mounting phase} or \textit{Idle phase} because nodes do not perform sampling during this phase.
This phase is the most laborious and time-consuming step of the process. 
Each sensor has a short wire with a connector attached at the end. 
Engineers at Space Systems Loral (SSL) have informed us that depending on sensor placements, partial disassembly of the satellite may occur to facilitate the sensor's placement in the exact locations needed to collect data.
From there, longer cables are run from the sensors to the \gls{DAQ}.
The mounting process can take between one to two weeks to attach the devices before the test is started. 
The cables have to be carefully mounted to the spacecraft for two primary reasons: avoid damage to the spacecraft during the test, and prevent significant changes to the spacecraft's load properties.
The latter is critical because if one wants to add more sensors to study the system better, the added weight may alter the system's load properties, which is in direct opposition to the test's goal.
This is because the more sensors added, the more likely the results are not representative of the realistic properties of the spacecraft~\cite{whatzwrong}.
Given that each cable performs slightly differently and has a unique transfer function, an individual {profile} needs to be generated so that these differences can be calibrated out. 
This wired solution is costly and lacks versatility.

After the mounting phase, each sensor has to be manually verified.
Verification is critical because engineers need to ensure that all sensors are functioning properly before the test is conducted. 
After verification, the tests begin.
This process is referred to as the \textit{Sampling phase}.
During the test, sensors are sampled continuously for 20 to 40 minutes.


\section{System Overview}
\label{sys_overview}
The system's three primary components are \textit{nodes}, \textit{\gls{AP}}, and a {\textit{server}}.
Each node is a wireless sensor device capable of sampling, processing, and wireless communication.
\gls{AP} receives packets from nodes and forwards them to the server.
The server coordinates network operation, decodes the received packets, processes the collected data, and generates time synchronization packets.

The basic components of a node are sensor (an accelerometer in our node design), \gls{ADC}, processor, and wireless transceiver.
The \gls{ADC} used in our node design is capable of sampling at up to $\textrm{500\:ksps}$ and is interfaced with the processor using \gls{SPI} bus.
The \gls{SoC} is based on CYW54907~\cite{CYW54907}, which supports 802.11ac.
This \glspl{SoC} includes two ARM-Cortex R4 processors, one processor is dedicated to the \textit{application subsystem}, and the other one is assigned to the \textit{wireless subsystem}. 
The former subsystem runs user code and the latter runs 802.11 firmware to handle time-critical tasks such as backoff and acknowledgment packet generation.
We discuss hardware design in~\cref{hardware_design}.

Considering the complexities of 802.11 standard, in this paper, we show that using various power saving modes are necessary to facilitate system development for a wide range of applications.
We refer to these power saving modes as \textit{Idle} and \textit{Alert}, where the former achieves higher energy efficiency while its responsiveness to incoming messages is slower than that of the latter.
We will explain these in \cref{section:idle_alert_phase}.

\begin{figure}[t]
\centering
\includegraphics[width=1\linewidth]{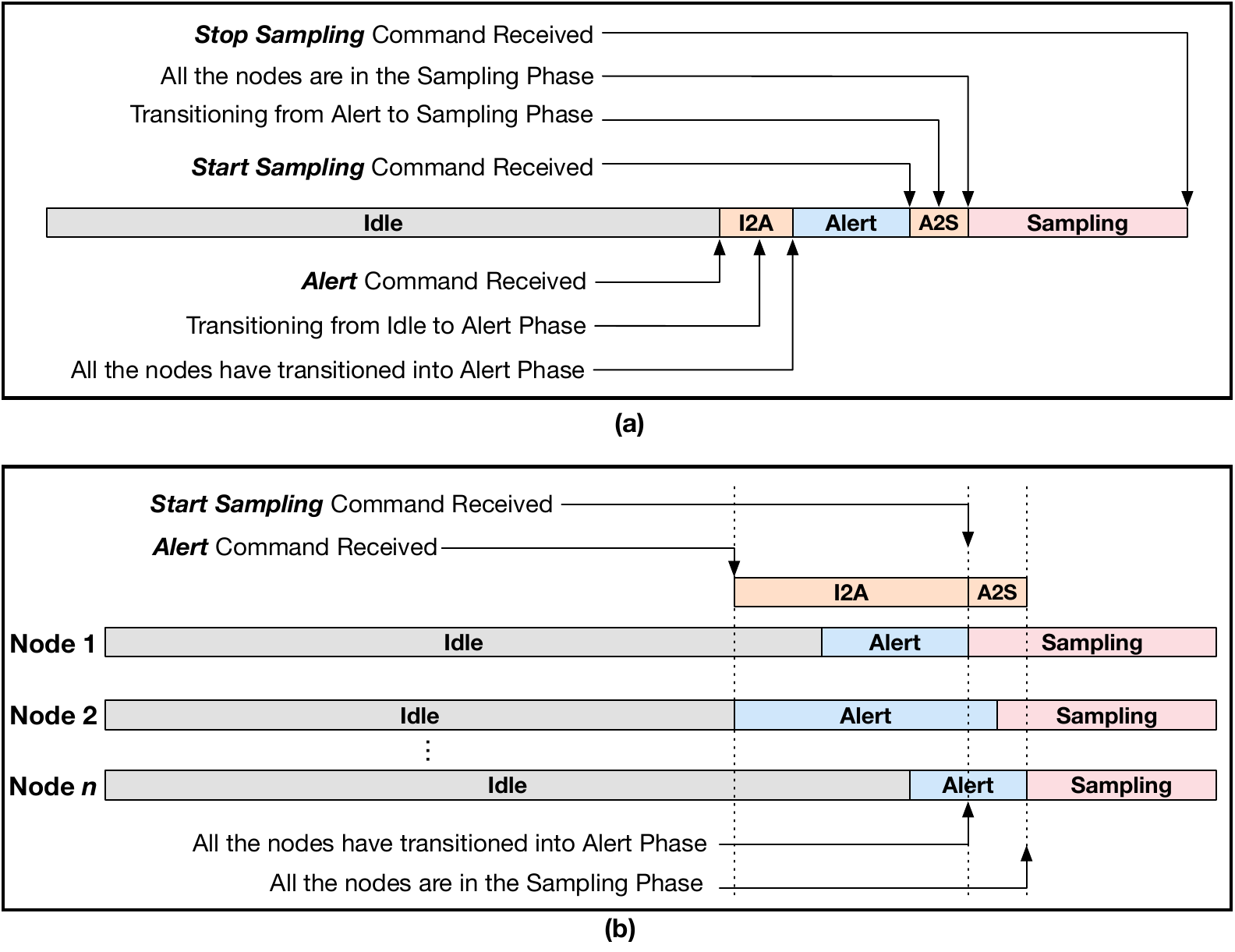}
\caption{(a) The operational phases of the overall system. 
(b) The operational phases of nodes.
Since the transition time of nodes varies depending on energy efficiency configuration, transition periods exist from the overall system's perspective.
These transition periods are named Idle to Alert (I2A) and Alert to Sampling (A2S).
}
\label{fig:op_phases}
\end{figure}

With regard to the vibration monitoring application (\cref{app_scenario}), Figure~\ref{fig:op_phases} presents the operational phases of the system.
We detail Figure~\ref{fig:op_phases}(a) as follows.
During the mounting phase, which may last between one to two weeks, nodes are mounted on the spacecraft.
Before mounting each node, the node is powered on and enters the Idle phase.
Once a node receives the \textit{Alert Command} from the server, the node transitions into the Alert phase and waits to receive a \textit{Start Sampling} command from the server to start the \textit{Sampling} phase.
Figure~\ref{fig:op_phases}(b) shows this transition for three nodes.
Since the nodes transition from the Idle phase to the Alert phase at different times, there is a transition period for the system to ensure all the nodes are in the Alert phase.
This delay is referred to as the \gls{I2A} transition.
Similarly, there exist an \gls{A2S} transition period.
Depending on the energy efficiency mode employed (\cref{section:idle_alert_phase}), \gls{I2A} duration varies between a few seconds to a few minutes; whereas, \gls{A2S} varies between a few hundreds of milliseconds to a few seconds.

The different energy efficiency methods used during the Idle phase and Alert phase justify the need for a low-cost transition from the Idle phase to the Sampling phase.
Specifically, if we send a Start Sampling command to the nodes in the Idle phase, they start sampling at time instances that may be up to a few minutes apart. 
Therefore, while some nodes are consuming higher energy to collect and transmit samples, the rest are still in low-power mode; and this results in partial system monitoring and energy waste by some nodes.
Considering the high energy consumption of nodes during the Sampling phase, we need to ensure all the nodes start the Sampling phase almost synchronously, however, this is not possible considering the deep sleep mode used during the Idle phase.
Therefore, we use an intermediate energy efficiency method---the Alert phase---to minimize the intervals between the instances the nodes transition into the Sampling phase.

\begin{figure}[t]
\centering
\includegraphics[width=1\linewidth]{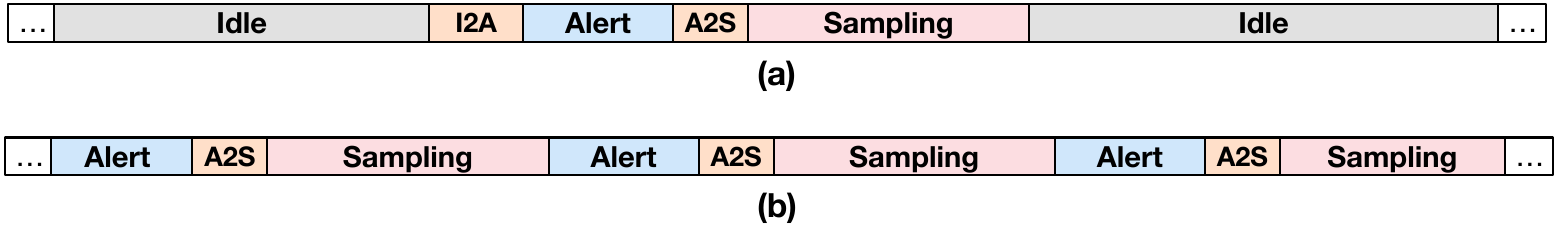}
\caption{(a) A system with long intervals between Sampling phases.
(b) A system with short intervals between Sampling phases.
}
\label{fig:op_phases_alternative}
\end{figure}

The Idle and Alert phases can be used to develop various 802.11-based \glspl{WSN}.
For example, for an application that requires periodical sampling of a structure at long intervals, the above three phases can be used, as Figure~\ref{fig:op_phases_alternative}(a) shows.
For applications where the interval between Sampling phases is short, only the Alert phase can be used to enhance the energy efficiency of the system, as Figure~\ref{fig:op_phases_alternative}(b) shows. 

Algorithm~\ref{alg:main_thread} presents the operation of a node during the Sampling phase.
\begin{algorithm}[t]
\footnotesize
\SetInd{0.5em}{0.5em}
\SetKwInOut{Input}{inputs}
\SetKwFunction{Fmain}{main\_thread}
\SetKwProg{Fn}{function}{}{}
\BlankLine
\Fn{\Fmain{}}
{  \BlankLine
    
    $n=0$ {\color{Blue}\tcc{initialize packet sequence number}}
   \BlankLine
   \While{sampling\_phase}
   {
       $n = n+1$ 
       
       \label{start_sampling}
       \For{$i=0\:;i<s\:;i++$}
       {{\color{Blue}\tcc{$s$ is the maximum number of samples per packet. We refer to $s$ as sampling batch size}}
          
          \texttt{$T[i]\:=\:\mathrm{\texttt{get\_nanosec\_clock()}}$}

          $S[i]\:=\:\mathrm{\texttt{get\_ADC\_sample()}}$ {\color{Blue}\tcc{2 bytes}}
          
            \uIf{$i > 0$}
            { 
                $I[i] = T[i] - T[i-1]$ {\color{Blue}\tcc{calculate inter-sample interval}}
            }
       }
       \label{end_sampling}
       
       \BlankLine
       \texttt{intv\_freq\_count($I[s],H[s],used$)} {\color{Blue}\tcc{calculate repetitions per inter-sample interval}}
       \label{start_encoding}
       
       \texttt{intv\_freq\_ranking($H[s],used$)}{\color{Blue}\tcc{rank inter-sample intervals based on repetitions}}
       \label{lst:line:ranking}
       \texttt{packet\_encoding($H[s],count,S[s],I[s],pkt$)}
       \label{end_encoding}
       
       \BlankLine

       \texttt{packet\_preparation($pkt$)} 
       \label{prepare_packet}       
       
       \texttt{packet\_send($pkt$)}
       \label{send_packet}              
   }
    \KwRet
}
\caption{The Overall Control Flow of Each Node During the Sampling Phase}
\label{alg:main_thread}
\end{algorithm}
Each node performs sampling, encoding, and packet transmission sequentially.
These operations repeat until the Stop Sampling command is received from the server.
The \textit{sampling task} communicates with the ADC to collect $s$ samples (lines~\ref{start_sampling} to~\ref{end_sampling}).
We refer to $s$ as the \textit{sampling batch size}.
The collected samples and timestamps are then encoded to reduce the overhead of timestamp assignment. 
This is achieved by first calculating repetitions per inter-sample interval (line \ref{start_encoding}), and then ranking inter-sample intervals based on repetitions (line \ref{lst:line:ranking}).
We detail packet encoding in \cref{sample_coll_and_enc}.
Next, the proper data structures are allocated to the packet (line~\ref{prepare_packet}), and then the packet is processed by the communication protocol stack (line~\ref{send_packet}).
We discuss packet creation and processing in~\cref{section:packet_processing}.

\section{Sampling and Packet Encoding}
\label{sample_coll_and_enc}

The server collecting samples must be able to accurately reconstruct the signals measured by the nodes.
If samples are collected at equidistant intervals, one timestamp per packet can specify the timestamps of all the samples included in the packet.
However, there are variations in sampling intervals; hence, including one timestamp per packet reduces the timing accuracy of the measurement event. 
On the other hand, assigning a per-sample timestamp introduces a significant overhead.
%

In this section, we propose methods for efficient and lossless compression of timestamps. 
Although these methods are application-agnostic, they are tailored for the packet size limitation of 802.11.
In addition to \textit{compression efficiency}, we also consider \textit{execution time} as an important design metric to minimize the effect of encoding on sampling rate.

\subsection{Sampling Interval Variations}
\label{sampling_intv}
\begin{figure}[!t]
\centering
  \includegraphics[width=1\linewidth]{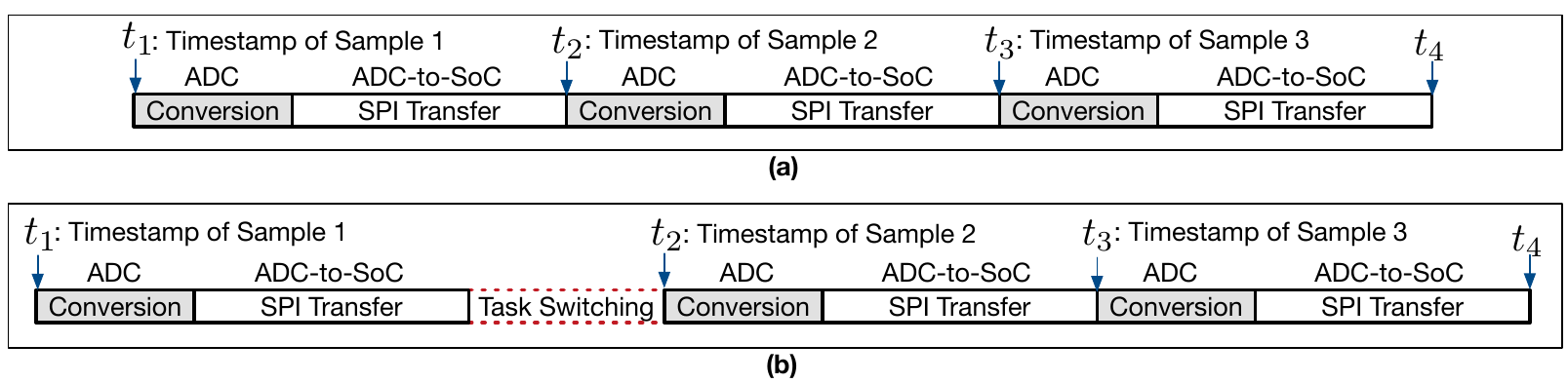}
  \caption{Sample collection from ADC.
  Both the conversion time of ADC and sample transfer time affect inter-sample intervals.
  Also, further variations of inter-sample intervals occur when the processor switches between tasks.
  }
  \label{fig:inter_sample_interval}
\end{figure}

In this section, we focus on inter-sample intervals from two points of view: \gls{ADC}'s sample conversion time and the delay of transferring a sample from the ADC to the processor.
The design of \gls{SAR} \glspl{ADC} includes an internal switched-capacitor \gls{DAC}, also known as charge redistribution \gls{DAC}.
The \gls{DAC} uses capacitors to store the voltage at the input and uses a switch to disconnect the capacitors from the input~\cite{zhai2018sample}.
The switched-capacitor structure's conversion time scales exponentially with the resolution of \gls{ADC}. 
Besides resolution, the amount of time to hold the capacitor value, impedance of the internal analog multiplexer, output impedance of the analog source, and the switch impedance are the factors impacting \gls{ADC}'s conversion time.
In addition to conversion time, the amount of time required to transfer a sample from the \gls{ADC} to the processor may vary.
Figure \ref{fig:inter_sample_interval}(a) presents the process of sample collection from ADC.
At the time $t_{1}$, and right before collecting the first sample, the processor collects the timestamp for Sample 1.
The processor then triggers a pin of the ADC, at which point the ADC starts the conversion.
Once the ADC finishes the conversion, the SPI communication starts to transfer the converted sample from the ADC to the SoC.

\begin{figure}[!t]
\centering 
 \subfloat[\scriptsize $\textrm{100\:ksps}$ at $\textrm{25\:}\mathrm{^{\circ}C}$\label{1b}]{
 \includegraphics[width=0.48\linewidth]{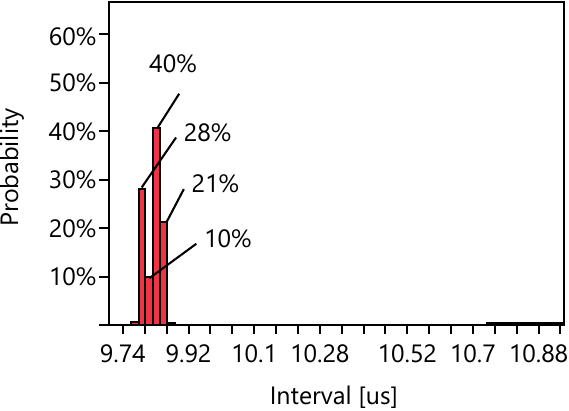}
 \hfill
} 
\subfloat[\scriptsize $\textrm{500\:ksps}$ at $\textrm{25\:}\mathrm{^{\circ}C}$\label{1.5b}]{
 \includegraphics[width=0.48\linewidth]{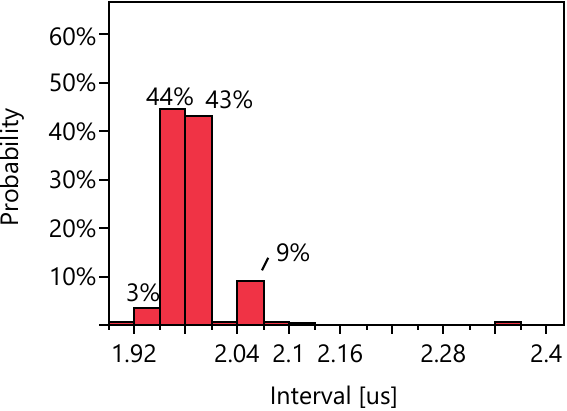}
 \hfill
} 
  \caption{{Distribution of sampling interval under different sample frequency. 
  (a): Sampling rate $\textrm{100\:k\gls{sps}}$.
  (b): Sampling rate $\textrm{500\:k\gls{sps}}$.  
  For these experiments, the only task being run by the nodes is sampling.
  Nearly $\mathrm{99\%}$ of the intervals are within four to six classes, and the remaining $\textrm{1\%}$ of the intervals are distributed among several classes. 
  It is worth mentioning that we observed similar distributions versus temperature variations.
  } 
  }
  \label{fig:interval_prob_100k}
 \end{figure}

To measure variations of inter-sample intervals (i.e., differences among $t_2 - t_{1}$, $t_3 - t_{2}$, etc.), we placed a node inside a controlled temperature chamber and used a high-accuracy logic analyzer to capture inter-sample intervals.
For these experiments, only the sampling task is run by the node; this means the processor runs a thread responsible for triggering the \gls{ADC} and transferring samples from the \gls{ADC} to the processor.
Figure~\ref{fig:interval_prob_100k} shows inter-sample intervals. 
For $\textrm{100\:ksps}$, we observe up to $\textrm{0.11\:}\mathrm{\mu}\textrm{s}$ variations around the mean value of $\textrm{10\:}\mathrm{\mu}\textrm{s}$, 
and for $\textrm{500\:ksps}$, the variations are up to $\textrm{0.016\:}\mathrm{\mu}\textrm{s}$ around the mean value of $\textrm{2\:}\mathrm{\mu}\textrm{s}$.
Considering the $\textrm{25\:ns}$ timing precision provided by the logic analyzer, nearly $\mathrm{99\%}$ of the intervals are within four to six classes for both sampling rates.
The other $\textrm{1\%}$ of the intervals are distributed among several classes.

%
Based on these observations, considering inter-sample variations is a crucial factor to improve timestamp encoding efficiency.
If we include only one timestamp per packet, these errors accumulate as more samples are included in the packet.
For example, when sampling at $\textrm{500\:ksps}$, for a packet including 500 samples, the error of the timestamp inferred for the last sample is about $\mathrm{8\:\mu s}$.
This error is about $\mathrm{55\:\mu s}$ for $\textrm{100\:ksps}$.

The inter-sample intervals reported in this section are irrespective of the load of the processor.
However, as Figure \ref{fig:inter_sample_interval}(b) shows, switching to other tasks exacerbates the variations of inter-sample intervals.
In \cref{pkt_proces_samplign_stability}, we will demonstrate and mitigate the effect of running other tasks (such as packet processing) on the variations of inter-sample intervals.
It is also important to note that the inter-sample variations reported in this section are pertaining to the specific hardware components used in our node design (\cref{hardware_design}).
Mitigation or elimination of inter-sample variations via hardware design is out of the scope of this work and is left as future work.


\subsection{Encoding Methods}
\label{encoding_methods}

\subsubsection{Baseline}
Our baseline method refers to assigning a timestamp per sample without applying any encoding.
We discuss two baseline methods as follows.

\begin{figure}[!t]
\centering
  \includegraphics[width=1\linewidth]{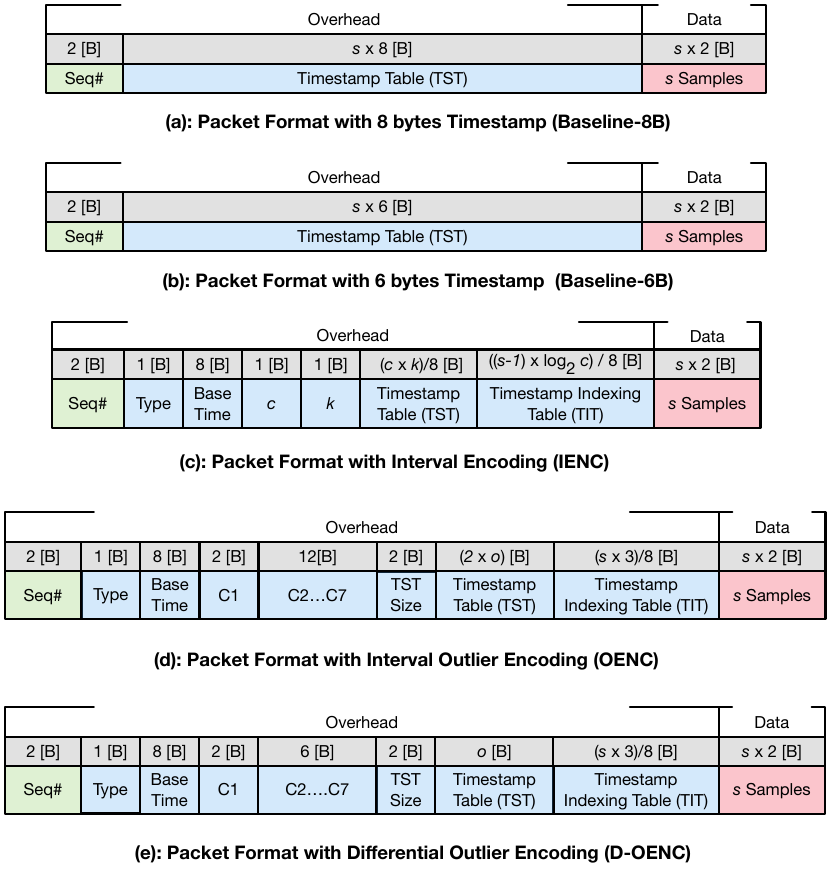}
  \caption{Packet formats of various timestamp encoding methods. 
  (a) Packet format with an $\textrm{8\:byte}$ timestamp per sample (Baseline-8B).
  (b) Packet format with an $\textrm{6\:byte}$ timestamp per sample (Baseline-6B).
  (c) Packet format using \acrfull{IENC}.
  (d) Packet format using \acrfull{OENC} where $\textrm{2\:bytes}$ are assigned to each outlier.
  (e) Packet format with \acrfull{D-OENC} where $\textrm{1\:byte}$ is assigned to each outlier.
  }
  \label{fig:packet}
\end{figure}

The \gls{SoC} provides a nanosecond timer that can be read into an $\textrm{8-byte}$ variable. 
Figure~\ref{fig:packet}(a) shows the baseline packet format where an $\textrm{8-byte}$ timestamp is added per sample.
Each sample is $\textrm{2\:bytes}$.
\textit{Seq\#} is the packet sequence number. 
Considering the 1472-byte \gls{MTU}, we can include a maximum of $\textrm{147\:samples}$ per packet, which means $\mathrm{80\%}$ of the payload is occupied by timestamps.
We refer to this method as \textit{Baseline-8B}, as demonstrated in Figure~\ref{fig:packet}(a).

The time duration provided by an $\textrm{8-byte}$, nanosecond-granularity variable is over 500 years, which is more than what is required in real-world applications.
Assuming time synchronization is only required during the Sampling phase, a $\textrm{60-minutes}$ Sampling duration requires $\textrm{42\:bits}$ to provide nanosecond timing accuracy.
Therefore, we use a $\textrm{6-byte}$ nanosecond timestamp, which wraps around every $\textrm{78\:hours}$.
With a $\textrm{6-byte}$ timestamp, we can include a maximum of 183 samples per packet, introducing a $\mathrm{75\%}$ overhead per packet for timestamps.
We refer to this method as \textit{Baseline-6B}, as demonstrated in Figure~\ref{fig:packet}(b).

To reduce the overhead of adding timestamps and increase the number of samples sent per packet, we present three lossless encoding methods and compare their performance against the baselines, as follows.


\subsubsection{\gls{IENC}}
The basic idea is to encode sampling interval classes into an index table.
The overhead of timestamp encoding varies based on the distribution of the interval classes.
Figure~\ref{fig:packet}(c) shows the packet format of this encoding method.
The two main tables used for timestamp encoding are \acrfull{TST} and \gls{TIT}.
The \gls{TST} includes the classes of sample conversion intervals, and each interval is encoded with $\textrm{2\:bytes}$ (instead of $\textrm{8\:bytes}$).
The total number of interval classes is denoted by the variable $c$.
Each interval is represented as a coefficient of $\textrm{1\:ns}$, and the number of bits required to represent each class is $k$.
Two variables $c$ and $k$ are used to inform the server about the size of the table to decode.
For each sample $S[i]$ where $i > 1$, there is a corresponding entry in the \gls{TIT} that refers to an entry of \gls{TST}.
Since \gls{TST} has $c$ entries, each entry in the \gls{TIT} is $\log_{2}c$ bits.
With this, the \gls{TIT} specifies the interval between each pair of consecutive samples.
\textit{Base Time} is the nanosecond timestamp of the first sample in the packet.

\begin{figure}[!t]
\centering 
\subfloat[\scriptsize 256-Samples Batch\label{2pa}]{
\includegraphics[width=0.475\linewidth]{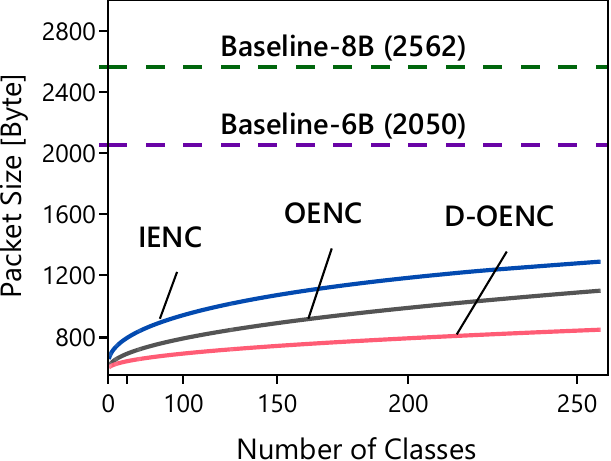}
\hfill
}
\subfloat[\scriptsize 512-Samples Batch\label{2pb}]{
\includegraphics[width=0.475\linewidth]{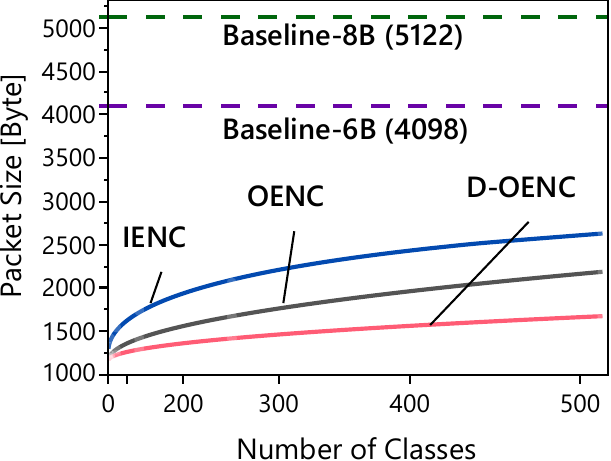}
\hfill
} 
\caption{Packet size comparison across the three encoding methods. 
X-axis represents the number of classes in a batch of samples ($s$).
Note that the x-axis increments by the power of 2.}
\label{fig:encoding}
\end{figure}

Figure~\ref{fig:encoding} shows the theoretical comparison of packet size using different encoding methods. 
The x-axis of sub-figures (a) and (b) represent the number of possible interval classes considering 256- and 512-sample batches, respectively.
Comparing with the baseline methods (horizontal lines), \gls{IENC} reduces payload overhead by at least $\mathrm{48\%}$ compared to {Baseline-8B} and $\mathrm{35.6\%}$ compared to {Baseline-6B}.

With 512-sample batches, \gls{IENC} hits the payload constraint when more than $\mathrm{12.5\%}$ of the intervals are unique. 
When reaching the payload limitation, \gls{IENC} stops processing the \gls{TST} and encodes the samples into two 256-sample packets. 
Notice that these two sub-divided packets have their own {Base Time}, \gls{TIT}, and \gls{TST} to ensure a packet loss would not influence the other one.
We use the \textit{Type} field to notify the server if this packet was sub-divided. 
The value 0 means only one packet, value 1 means the first of two packets, and value 2 represents the second of two packets.

To further improve encoding efficiency, we note that \gls{IENC} does not consider the repetitions of interval classes.
We rely on this observation and propose two encoding methods, as follows.

\subsubsection{\gls{OENC}}
This method leverages the statistical pattern of the classes to reduce encoding overhead.
Figure~\ref{fig:interval_prob_100k} shows that a few classes include the majority of the intervals. 
We refer to these classes as $C1$ through $C7$.
For example, in most 512-sample batches, only 6 intervals are outside of these major classes.
We use this observation to further reduce the \gls{TIT} size.
Figure~\ref{fig:packet}(d) shows the packet format used by \gls{OENC}.
The basic idea is to reduce \gls{TIT} size by using a $\textrm{3-bit}$ index for each sample.
This method uses a fixed \gls{TIT} size instead of letting the size grow by the number of interval classes. 
In the \gls{TIT}, there is a $\textrm{3\:bit}$ index for each sample; therefore the size of \gls{TIT} is $s \times 3\;\textrm{bits}$.
Index values 000 through 110 refer to $C1$ through $C7$.
If the index of a sample is 111, it means the interval is an outlier and must be found in the \gls{TST}.
For each index value 111 in the \gls{TIT}, there is a corresponding outlier value in the \gls{TST}.
In other words, the number of entries with index value 111 in the \gls{TIT} is the same as the number of entries in \gls{TST}.
The \textit{TST Size} field is used to inform the server about the size of \gls{TST}. 

Figure~\ref{fig:encoding} shows that \gls{OENC} reduces packet payload size by at least $\mathrm{56\%}$ compared to  \textit{Baseline-8B} and $\mathrm{45\%}$ compared to \textit{Baseline-6B}.
Compared to \gls{IENC}, the reduction is about $\mathrm{8\%}$ with the maximum number of outliers.
\gls{OENC} hits the payload constraint when more than $\mathrm{23\%}$ of the intervals in a 512-sample batch are unique.
Once it reaches the limit, \gls{OENC} stops processing the \gls{TST} and sub-divides the samples into two packets. 
These two packets create their own $C1$ to $C7$, \gls{TST}, \gls{TST} size, and the \gls{TIT} field values.
Figure~\ref{fig:encoding}(a) shows that \gls{OENC}'s worst case scenario with 256 samples consumes $\textrm{1132\:bytes}$, which is within the payload limit.
As Figure~\ref{fig:packet}(d) shows, \gls{TIT} always consumes $\textrm{96}$ and $\textrm{192\:bytes}$ of the payload for 256- and 512-sample batches, respectively; whereas, \gls{TST} grows as the variations of sampling intervals increase. 
In the worst case, \gls{OENC}'s \gls{TST} field occupies about $\mathrm{46\%}$ of the payload, which is the second-largest payload consumption field after the samples field.

\subsubsection{\gls{D-OENC}} 
This method uses lossless compression of \gls{TST} to reduce the number of bits required to encode each outlier.
Instead of encoding actual interval values, we encode a signed number to indicate the difference between each outlier and the first majority class, i.e., $C1$.
In other words, we encode the difference between the expected value and perceived value.
This signed number represents the difference as a multiple of clock accuracy (one nanosecond).
For example, if an inter-sample interval $I[i]$ is an outlier, the \gls{TST} entry corresponding to this outlier is $I[i] - I[C1]$, where $I[C1]$ is the inter-sample interval of class $C1$. 
Figure~\ref{fig:packet}(e) presents the packet format.

Regarding performance, Figure~\ref{fig:encoding} shows that \gls{D-OENC}'s packet payload size is reduced by at least $\mathrm{66\%}$ compared to \textit{Baseline-8B} and $\mathrm{57\%}$ compared to \textit{Baseline-6B}.
This is a $\mathrm{9\%}$ reduction compared with \gls{OENC} when all the outliers are unique.
In addition, \gls{D-OENC} hits payload constraint when more than $\mathrm{47\%}$ of the intervals are unique with 512-sample batches.
Once it reaches the limitation, \gls{D-OENC} simply splits the 512 samples into two packets.
As Figure~\ref{fig:encoding}(b) shows, \gls{D-OENC}'s worst-case scenario with a 512-sample batch consumes $\textrm{1741\:bytes}$. 
When split into two packets, each packet consumes $\textrm{1229\:bytes}$, well within the payload limit.


\subsection{Interval Classification Algorithm}
\begin{algorithm}[t]
\footnotesize
\SetInd{0.5em}{0.5em}
\SetKwFunction{Fmain}{intv\_freq\_count}
\SetKwProg{Fn}{function}{}{}
\BlankLine
\Fn{\Fmain{$I[s],H[s],used$}}
{
   \BlankLine
    $used=0$ {\color{Blue}\tcc{consumed array space}}
    
    $packet\_count = 1$ {\color{Blue}\tcc{number of packets}}
    
    \BlankLine
    {\color{Blue}\tcc{generate interval frequency hash table}}
    \For{$i=0\:;i<s\:;i++$} 
    {   
       {\color{Blue} } 
       \uIf{ $used < payload\_limit$}
       {\label{lst:line:payload_limit}
       
       \For{$j=0\:;j<s\:;j++$}
       {  
          \uIf{$i==0$}
            {  
               $H[used].value=I[i]$
               
                $H[used].count=1$
                
                break
            }
            \uElseIf{$j > used$}{
            
                $used++$
                
                $H[used].value=I[i]$
                
                $H[used].count=1$
                
                break
            }
            \uElse{
                \uIf{$H[j]==I[i]$}
                {  
                   $H[j].count++$
                   
                   
                   break
                }

            }
       }
      
    }
    \uElse{
       $packet\_count=2$
       
       break
         }
    }
    
    {\color{Blue}\tcc{only for \gls{IENC} and \gls{OENC}}}
    \uIf{$packet\_count ==2$}
    {
        generate the interval frequency hash table with $s/2$ 
    }
    
   \KwRet $H[s]$,$used$ 
}
\BlankLine

\caption{Interval Classification Algorithm}
\label{alg:interval_class_algo}
\end{algorithm}

We use the hash table as a simple and effective method to count the frequency of sampling intervals, as demonstrated in Algorithm~\ref{alg:interval_class_algo}.
In our implementation, we only use stack memory to avoid the computation time uncertainty of dynamic memory.
The variable $used$ keeps track of the consumed array space by the number of classes. 
Line~\ref{lst:line:payload_limit} shows that when the number of classes reaches the $payload\_limit$, the algorithm breaks the loop and recomputes the hash table with half of the samples. 
This recomputing process happens with \gls{IENC} and \gls{OENC} only. 
After generating the frequency hash table $H[s]$, we use this table to compute the probability distribution of sampling intervals. 
Function \texttt{intv\_freq\_ranking()} is used to find the top seven classes---$C1$ through $C7$. 
The function computation time is based on the size of the consumed portion of $H[s]$, which is $O(used)$.
Only \gls{OENC} and \gls{D-OENC} use this function to rank the top seven classes.


\subsection{Execution Time}
\label{encoding_exec_time}
\begin{figure}[!t]
\centering 
\subfloat[\scriptsize 256-Sample Batch\label{ex_a}]{
\includegraphics[width=0.475\linewidth]{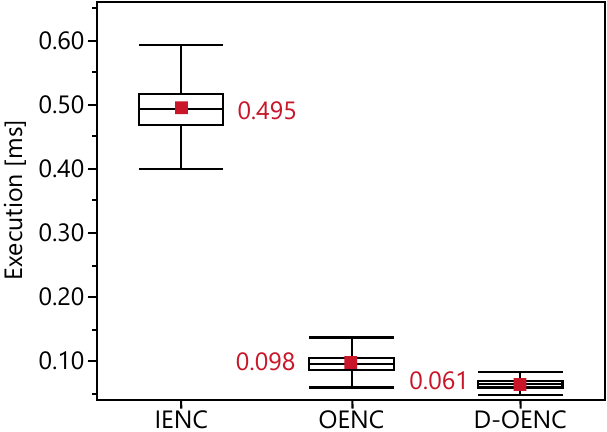}
\hfill
}
\subfloat[\scriptsize 512-Sample Batch\label{ex_b}]{
\includegraphics[width=0.475\linewidth]{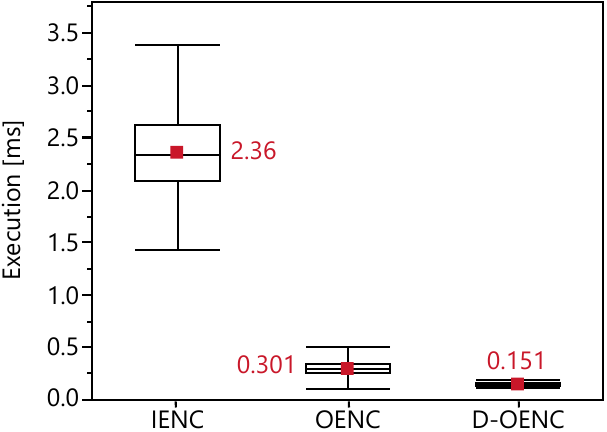}
\hfill
} 
\caption{Empirical evaluation of the execution time of the proposed encoding methods. 
The processor is an ARM-Cortex R4 operating at 160 MHz.
The numbers next to box plots represent mean values.
\gls{IENC} shows the largest variations because the chance of breaking a batch of samples into two packets is the highest with this encoding method.
\gls{D-OENC} achieves the minimum execution time among the three methods.
}
\label{fig:execution}
\end{figure}

Figure~\ref{fig:execution} shows the empirical evaluation of execution time. 
%
The \gls{IENC} computation time is more than 5x and 7x longer than \gls{OENC} and \gls{D-OENC} when using 256- and 512-sample batches, respectively.
This is mainly because the number of interval classes impacts \gls{IENC}'s computation time linearly, causing these large variations.
Using 512-sample batches, \gls{IENC} hits the packet payload limit when more than $\mathrm{12.5\%}$ of the intervals are unique (64 interval classes).
This situation results in encoding the samples into two packets, which requires \gls{IENC} to recompute the hash table.
The \gls{OENC} method performs a similar operation when more than $\mathrm{23\%}$ of the intervals are unique.
However, \gls{D-OENC} does not need to recompute the hash table when splitting into two packets is required.

As Algorithm \ref{alg:main_thread} shows, during the Sampling phase, each node sequentially performs sampling, encoding, packet preparation, and transmission.
Therefore, a shorter encoding period is essential to achieve a stable and higher sampling rate.
When the sampling rate is $\textrm{100\:k\gls{sps}}$, these results show that \gls{IENC}, \gls{OENC}, and \gls{D-OENC} result in missing about 49, 9, and 6 samples after collecting each 256-sample batch.
When the batch size is 512, these values are 236, 30, and 15 samples.
In summary, the results presented in Figures \ref{fig:encoding} and \ref{fig:execution} confirm the superiority of \gls{D-OENC} in terms of packet size efficiency and execution duration.
When the batch size increases from 256 to 512, the execution time of \gls{D-OENC} shows a $\mathrm{147\%}$ increase.
Although using the smaller batch seems to be more efficient, we show in \cref{perf_eval} that using the larger batch results in higher sampling efficiency.

\subsection{Compression Ratio}
\label{compression_ratio_eval}
\begin{figure}[!t]
\centering 
\subfloat[\scriptsize 256-Samples Batch\label{2com}]{
\includegraphics[width=1\linewidth]{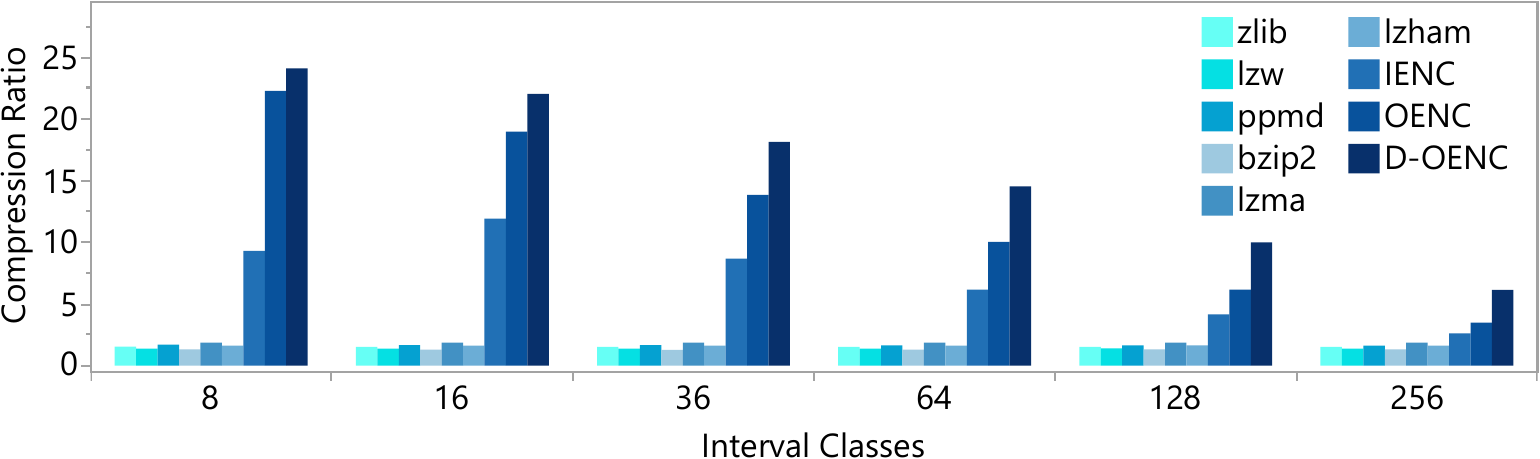}
\hfill
}
\\
\subfloat[\scriptsize 512-Samples Batch\label{2com}]{
\includegraphics[width=1\linewidth]{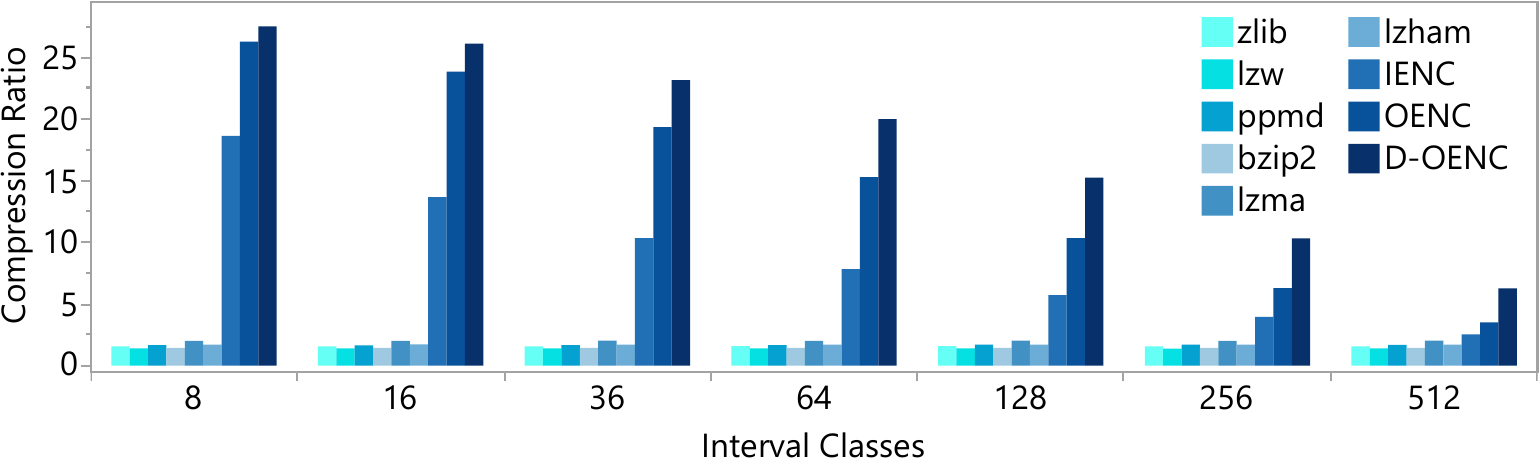}
\hfill
} 
\caption{Comparing the compression ratio of the proposed algorithms versus standard lossless compression algorithms. The x-axes represent the number of the timestamp interval classes in a batch of (a) 256- and (b) 512-samples.}
\label{fig:compression}
\end{figure}

This section compares the proposed algorithms versus a set of well-known lossless algorithms that use the static dictionary and adaptive dictionary techniques.
Specifically, we consider the followings:
\begin{itemize}
    \item Lempel-Ziv algorithms: Sliding Window LZ77 (\textit{zlib}) and Dictionary Based LZ78 (\textit{lzw})~\cite{barr2006energy},
    \item Markov modeling following by arithmetic coding: Prediction with Partial Match (\textit{ppmd})~\cite{barr2006energy},
    \item Burrows-Wheeler Transform (bzip2)~\cite{pankratius2009parallelizing}, 
    \item Lempel–Ziv–Markov chain algorithm: standard (\textit{lzma}) and advanced (\textit{lzham})~\cite{habib2019dictionary}.
\end{itemize}
The datasets used in examining the algorithms include timestamps, and we consider a various number of classes in 256- and 512-sample batches.
Figure~\ref{fig:compression} presents the empirical evaluation of the compression ratio.
One of the unique features of the proposed algorithms is to encode the non-repetitiveness of timestamps as inter-sample intervals.
And for \gls{D-OENC}, it encodes the differences between expected and realized intervals between timestamps.
The results demonstrate that the proposed algorithms benefit from encoding intervals and achieve a higher compression ratio with fewer classes.
Because of the non-repetitiveness of timestamps, the standard lossless algorithms can only achieve a maximum compression ratio of 1.87 and 2.02 for 256- and 512-sample batches, respectively.
Whereas, the proposed algorithms can achieve compression ratios up to 24 and 27 for 256- and 512-sample batches.
Even with cases where all the sample intervals are unique, the proposed \gls{D-OENC} method achieves a maximum compression ratio of 6.15 and 6.27 for 256- and 512-sample batches, respectively.  

\section{Reducing Energy Consumption During the Idle and Alert Phases}
\label{section:idle_alert_phase}
Achieving energy-delay tradeoffs with the 802.11 standard is more complicated than low-power standards such as 802.15.4 and 802.15.1.
This is specifically because 802.11 stations need to periodically receive the beacons sent by the \gls{AP}, or reassociate with the \gls{AP} when they need to communicate.
Considering these challenges, we propose two methods for establishing energy-delay tradeoffs and facilitating the development of various 802.11-based applications, regardless of the hardware platform used.


\subsection{Energy Efficiency Methods}
We detail two energy efficiency methods in this section: 
\textit{periodic association}, and \textit{extended-period beacon reception}.

\subsubsection{Periodic Association}
In this method, each node periodically wakes up and associates with the \gls{AP}, checks for a pending command from the server, and then either returns to an off mode (if no pending command) or transitions into the requested mode.
The periodic association is triggered by a timer that fires every $t_{p}$ seconds to perform the association.
The energy consumed per second is computed as follows:
\begin{equation}
  \begin{gathered}
    \bar{E} = \bigg(P_{asc} \times D_{asc} + P_{off} \times ( t_{p} - D_{asc})\bigg)/t_{p} \\
    \mathrm{for}\;\; D_{asc} < t_{p}
    \label{pa_energy}
  \end{gathered}
\end{equation}
where $P_{asc}$ refers to the average power consumed during the association process, and $D_{asc}$ is the duration of the association process (the time it takes for the node to associate with an \gls{AP}).
$P_{off}$ is power consumption during the off mode.

Although a node's energy consumption can be reduced by increasing $t_{p}$, it is also possible to enhance energy efficiency by improving the association process.
We utilize two methods to reduce this energy.
(\romannumeral 1) \textit{Specific Association}:
During the association process, a node sends probe packets on all channels to discover nearby \glspl{AP}.
To avoid this overhead, we program the MAC address and channel of the \gls{AP} into the node's driver.
This allows each node to directly send an association request to the intended \gls{AP}, thereby reducing probing overhead.
Resilience against interference and association failures can be provided by adding a channel scan method to the nodes' embedded software.
Specifically, suppose a node fails to connect to the \gls{AP} over the programmed channel.
In that case, it switches to the scan mode to determine the new operational channel of the \gls{AP}. 
This new channel is then programmed into the driver to be used in subsequent associations.
(\romannumeral 2) \textit{\gls{PMK} Offloading}:
During the association process between a node and the \gls{AP}, the node needs to establish a secure channel with the \gls{AP}, and this entails generating a \gls{PMK}, which is a process-intensive task.
Using the driver's APIs, we offload this key generation process to the application subsystem and then transfer the key to the wireless subsystem.

\subsubsection{Extended-Period Beacon Reception}
In this method, as soon as a node is turned on, it is associated with the \gls{AP}.
To ensure low-power operation while maintaining association, we employ \gls{PSM}, which is the fundamental power saving mechanism supported by the 802.11 standard.
With \gls{PSM}, an associated node wakes up every $\textrm{102.4\:ms}$ to receive a beacon packet from the \gls{AP}.
When there is a buffered packet for a node, the \gls{AP} sets a flag (per node) in the beacon packet to inform the node that it must request for the delivery of the buffered packet.
Although the standard wake-up interval used by commercial \glspl{AP} is $\textrm{102.4\:ms}$, we can use a longer interval to reduce the overhead of beacon reception further.
We refer to this method as \textit{extended-period beacon reception}.
To this end, we modified the \gls{PSM} configuration in the driver (of nodes) to extend the default listen interval by a factor of $l$. 
We refer to $l$ as the \textit{listen interval coefficient}.
With this method, the energy consumed per second is calculated as follows:

\begin{multline}
    \bar{E} = P_{bcn}\times \left (  \frac{D_{bcn}}{l \times 0.1024}  \right ) + P_{slp}\times \left ( 1 - \frac{D_{bcn}}{l \times 0.1024}  \right )
    \label{pbr_energy}
\end{multline}

where $P_{bcn}$ and $P_{slp}$ are power consumption during beacon reception and sleep modes, respectively, $D_{bcn}$ is the duration of a beacon reception, and $l$ is the coefficient of the listen interval used by the node.
Note that $D_{bcn}$ includes the duration a node waits in receive mode to receive a beacon, as well as the actual duration of beacon packet reception.

\subsection{Transition Delay}

Using the periodic association method, although a larger value of $t_{p}$ results in a deeper sleep mode, the larger value also affects the nodes' responsiveness to commands. 
Assuming the command packet sent by the server can be generated at any time, the delay of command reception by a node is a uniform random variable in the range $[0, t_{p}]$ second.
Using the extended-period beacon reception method, the delay of receiving the command is a uniform random variable in the range $[0,  l\times 0.1024 ]$ second.
We introduce a definition of transition delay that makes it independent of command packet generation instance:
\textit{transition delay is defined as the interval between the time instances the first and last node in the network receive a command packet}.

\begin{figure}[!t]
\centering
  \includegraphics[width=1\linewidth]{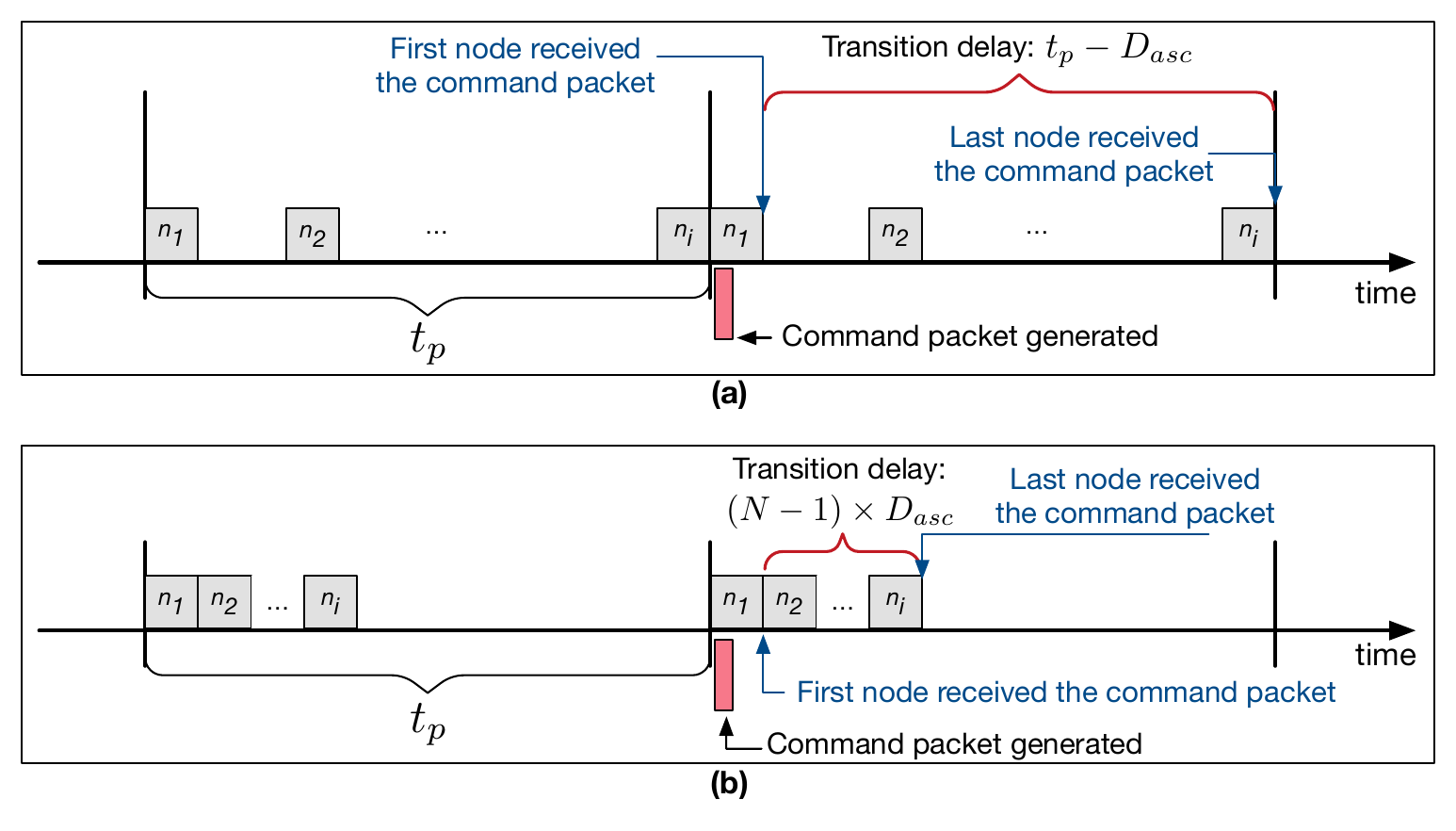}
  \vspace{-20pt}
  \caption{Transition delay of periodic association method.
  (a) Without applying time synchronization, the transition period is $t_{p} - D_{asc}$.
  (b) By coordinating the association time of nodes, the transition period is reduced to $(N-1) \times D_{asc}$ in a network including $N$ nodes.
  }
  \label{fig:trans_delay_per_asc}
\end{figure}

When using the periodic association method, Figure~\ref{fig:trans_delay_per_asc}(a) shows that the transition delay is $t_{p} - D_{asc}$.
For periodic association, the wake-up instance of the nodes can be coordinated if they are time-synchronized.
Although it is desirable to allow all the nodes to associate with the \gls{AP} concurrently, this results in channel access contention, extends the association duration, and increases the nodes' energy consumption.
An alternative approach is to time synchronize the nodes and serialize their associations, as Figure~\ref{fig:trans_delay_per_asc}(b) demonstrates.
For a system including $N$ nodes, this method reduces transition delay to $(N-1)\times D_{asc}$.
For this method, a clock accuracy of up to a few tens of milliseconds would suffice.
Each node synchronizes its clock with the server at each association instance.

\begin{figure}[!t]
\centering
  \includegraphics[width=1\linewidth]{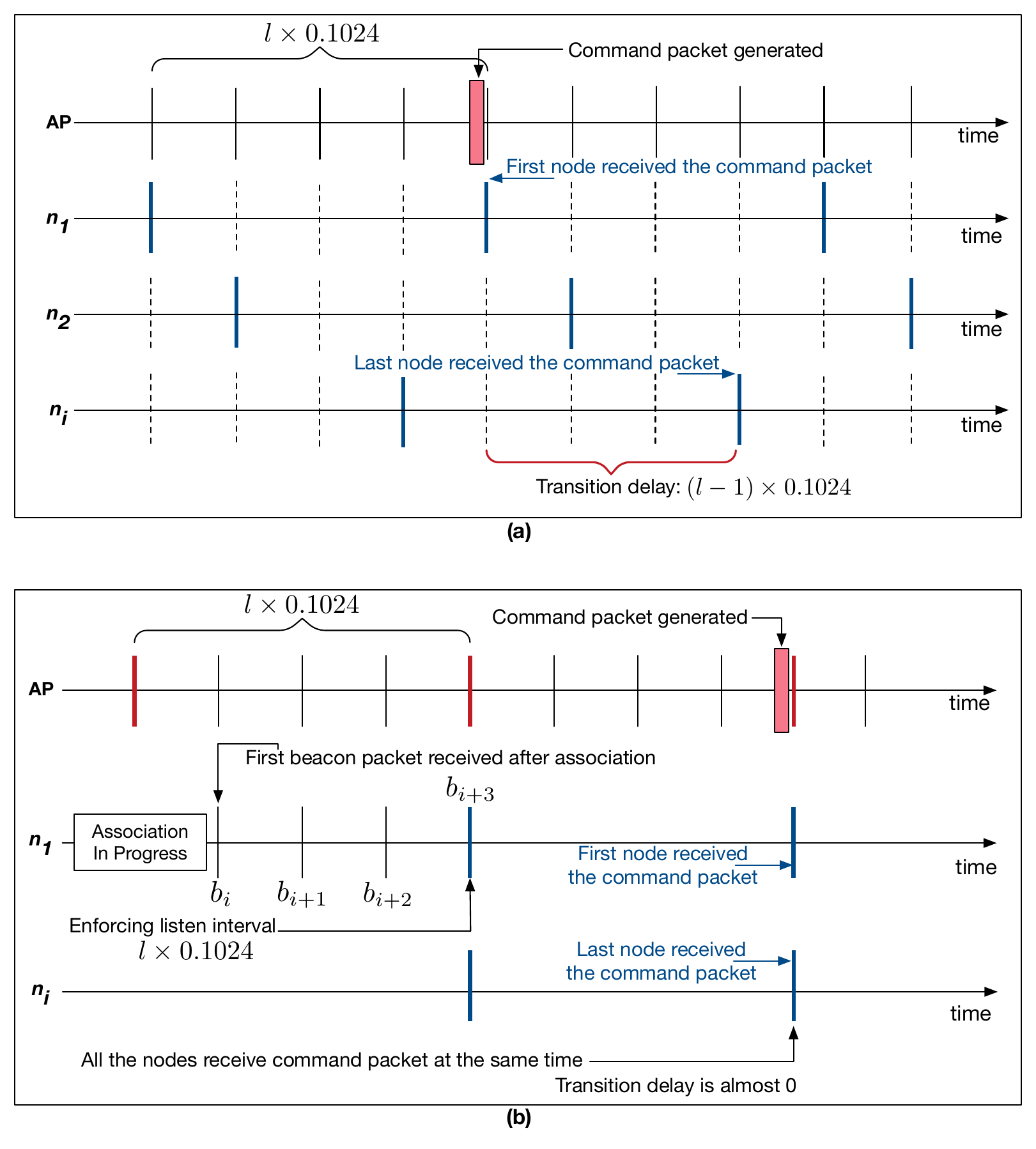}
  \vspace{-20pt}
  \caption{Transition delay of extended-period beacon reception method.
  (a) Without synchronizing the wake-up instance of nodes, transition duration is $(l-1)\times 0.1024$.
  (b) Transition delay reduces to almost zero by coordinating the wake-up time of all the nodes.
  }
  \label{fig:trans_delay_beacon_rec}
\end{figure}

For the extended-period beacon reception method, the server can issue the command right before all the nodes' wake-up instance.
However, although each node wakes up every $l \times 0.1024$ second, their wake-up instances are not synchronized; therefore, transition delay equals $(l-1)\times 0.1024$.
This is demonstrated in Figure~\ref{fig:trans_delay_beacon_rec}(a).
We employ the following method to tackle this challenge.
Once each node associates with the \gls{AP}, the server informs the node about the sequence number of the initial beacon ($b_{init}$) packet, where $b_{i} = b_{init} + k \times l, k \in \mathbb{N}$ represents the beacon instances that must be received by all the nodes.
Assume a node associates with the \gls{AP} and receives beacon number $b_{i}$.
If $b_{i} = b_{init} + k \times l$, the node immediately enforces listen interval coefficient $l$.
Otherwise, the node uses $l=1$ until $\mathrm{min}(b_{init} + k\times l)$, where $b_{i}< b_{init} + k\times l$ and $k \in \mathbb{N}$.
Then, the node starts enforcing the listen interval coefficient $l$.
For example, in Figure~\ref{fig:trans_delay_beacon_rec}(b), after the association of node $n_{1}$ is complete, this node keeps receiving beacons $b_{i}$ through $b_{i+3}$. 
Once beacon $b_{i+3}$ is received, the node enforces the listen interval coefficient $l$.
In the worst case scenario where a node completes its association right after receiving beacon $b_{i} = b_{init} + k \times l$, the node needs to receive $l$ beacons before enforcing listen interval.
With this method, the transition delay is negligible.

\subsection{Energy Comparison}
\label{idle_alert_energy}
\begin{figure}[!t]
\centering 
\subfloat[\scriptsize Periodic Association\label{5a}]{
\includegraphics[width=0.475\linewidth]{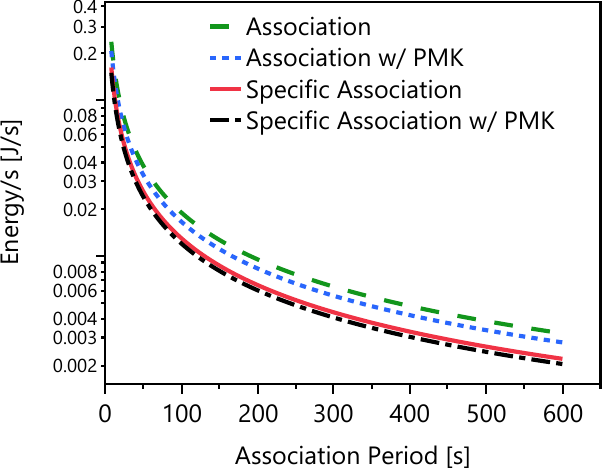}
\hfill
}
\subfloat[\scriptsize Extended-Period Beacon Reception\label{5b}]{
\includegraphics[width=0.475\linewidth]{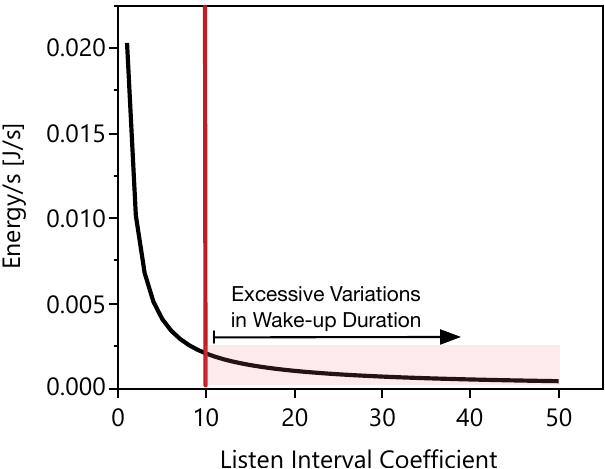}
\hfill
} 
\caption{Empirical energy evaluation of \gls{SoC}.
These results were collected in an interference-free environment.
}
\label{fig:energy_consumed}
\end{figure}

Figure~\ref{fig:energy_consumed} shows the empirical measurement of energy consumption per second by the \gls{SoC} (CYW54907) for the two proposed methods.
For these results, we first measured the duration and energy consumption of various operations, and then used Equations~\ref{pa_energy} and~\ref{pbr_energy}.
%
Since the energy consumed during the association process is high ($P_{asc} = \textrm{1.2\:J}$ for Specific Association w/PMK), a long association period ($t_{p}$) is required to achieve a reasonable level of energy efficiency.
With beacon listen interval coefficient $l=10$, the energy consumed per second is $\textrm{0.00205\:J}$, as Figure~\ref{fig:energy_consumed}(b) shows.
The same level of energy efficiency requires $t_{p}=593$ when using the Specific Association w/PMK method, confirmed by Figure~\ref{fig:energy_consumed}(a).

Although the extended-period beacon reception method seems to outperform the periodic association method, we noticed that as the listen interval coefficient increases, nodes become more sensitive to beacon loss.
Specifically, with a larger listen interval coefficient, a beacon loss results in longer energy consumed per beacon reception instance.
For example, we noticed that $l > 10$ triggers this behavior with CYW54907.
We tried different hardware platforms (BCM4343W, CYW43455, CYW43364) and observed that various transceivers demonstrate such behavior for different listen interval threshold levels.
For example, BCM4343W reveals this behavior for $l > 50$.
By investigating the 802.11 firmware, we observed that the link quality measurement algorithm requires receiving a certain number of beacons per unit time to measure link quality to the \gls{AP}.
Therefore, in environments where beacon loss may happen due to interference or low link quality, the listen interval coefficient must be carefully adjusted based on the characteristics of the 802.11 transceiver used.

It is worth noting that in this paper, we relied on the energy efficiency methods available by 802.11ac.
The subsequent of 802.11ac, known as \textit{802.11ax}, provides a new feature called \gls{TWT}, which allows stations to negotiate recurring wake-up instances with the \gls{AP}.
We leave the utilization of this method for achieving energy-latency trade-offs as future work.

\subsection{Energy Efficiency Parameters}
\label{energy_eff_parameters}
In this section, considering the energy efficiency methods given in \cref{section:idle_alert_phase}, we present models for configuring $t_{p}$ and $l$.
These models can be leveraged for the design of various 802.11-based applications.

Considering the scenario given in Figure \ref{fig:op_phases}, the following inequality must hold to satisfy the energy-efficiency requirement:
\begin{equation}
\begin{split}
   \label{eqn:energy_sum}
     \left. E_{idl}(t_{idl})+E_{i2a}(t_{i2a})+E_{alt}(t_{alt})+ E_{a2s}(t_{a2s}) \right.\\
     \left. +E_{smp}(t_{smp})  < E_{bat} \times 80\%  \right. \\
     \left. \mathrm{Subject\;to}: \: 1 \leq l \leq 10  \;\;\mathrm{and}\;\; D_{asc} < t_{p} \right.
\end{split}
\end{equation}
where $E_{idl}(\cdot)$, $E_{i2a}(\cdot)$, $E_{alt}(\cdot)$, $E_{a2s}(\cdot)$, and $E_{smp}(\cdot)$ are functions to compute the energy consumed during the Idle phase, \gls{I2A} transition, Alert phase, \gls{A2S} transition, and Sampling phase, respectively.
$E_{bat}$ is the total available energy of the battery, and we assume only $\mathrm{80\%}$ of the battery capacity can be used.
In the rest of this section, we assume $E_{a2s}(t_{a2s}) = 0$ as we showed in \cref{section:idle_alert_phase}.
Considering the energy efficiency models given in \cref{section:idle_alert_phase}, we extend Inequality \ref{eqn:energy_sum} as follows:

   \begin{align}
    \label{overall_energy_model}
   \left. \underbrace{t_{idl} \times \Bigg(\bigg(P_{asc} \times D_{asc}  +  P_{off} \times ( t_{p} - D_{asc})\bigg)/t_{p}\Bigg)}_\text{(a) Idle phase}  + \right. \nonumber \\
  \left. \underbrace{(l \times 0.1024) \times \Bigg(P_{bcn}  (  \frac{D_{bcn}}{0.1024}  ) + P_{slp}  ( 1 - \frac{D_{bcn}}{0.1024}  )\Bigg)}_\text{(b) Alert phase: Enforcing Listen Interval $l$}  + \right. \nonumber \\   
  \left. \underbrace{\Bigg((N-1) \times D_{asc} - (l \times 0.1024)\Bigg) \times \Bigg(P_{bcn} (  \frac{D_{bcn}}{l \times 0.1024} ) }_\text{(c) Alert phase: I2A transition of other nodes} + \right.\nonumber \\ 
  \left. \underbrace{ P_{slp}\times ( 1 - \frac{D_{bcn}}{l \times 0.1024}  )\Bigg)}_\text{(c) Alert phase: I2A transition of other nodes (continued)} + \right. \nonumber \\
    \left. \underbrace{(l \times 0.1024) \times \Bigg(P_{bcn}\times (  \frac{D_{bcn}}{l \times 0.1024} ) }_\text{(d) Alert phase: Enforcing Listen Interval $l$ by last node} + \right.\nonumber \\
    \left. \underbrace{P_{slp}\times ( 1 - \frac{D_{bcn}}{l \times 0.1024}  )\Bigg)}_\text{(d) Alert phase: Enforcing Listen Interval $l$ by last node (continued)} + \right. \nonumber \\
   \underbrace{t_{smp} \times \bar{E}_{smp}}_\text{(e) Sampling phase}    \nonumber \\
    \left. <  0.8 \times E_{bat}\right. \nonumber \\
    \left. \mathrm{Subject\;to}: \: 1 \leq l \leq 10  \;\;\mathrm{and}\;\; D_{asc} < t_{p} \right.    
   \end{align}

We detail this model as follows.
We assume the periodic association method is used during Idle phase, and the extended-period beacon reception is used during the Alert phase.
This is because the periodic association method can be used to achieve a deeper sleep mode.
$E_{idl}(t_{idl}) = t_{idl} \times \bar{E}_{idl}$, where $\bar{E}_{idl}$ is computed in Equation~\ref{pa_energy}, as demonstrated in Equation~\ref{overall_energy_model}(a).
Once the first node transitioned into the Alert phase, it takes $(N-1) \times D_{asc}$ seconds for the rest of the nodes to complete their transition.
During this transition duration, the energy consumed by the node can be computed using Equation~\ref{pbr_energy}.
However, note that the node may need up to $l \times 0.1024$ seconds before enforcing the listen interval $l$, as discussed in~\cref{section:idle_alert_phase}.
The energy consumed in this mode is computed as $(l \times 0.1024) \times \bar{E}_{alt}$, where $\bar{E}_{alt}$ is computed by Equation~\ref{pbr_energy}. 
Note that the value of $l$ in Equation~\ref{pbr_energy} must be 1.
This is demonstrated in Equation~\ref{overall_energy_model}(b).
After enforcing the listen interval $l$, the energy consumed during $(N-1)\times D_{asc} - (l \times 0.1024)$ is computed in a similar manner using Equation \ref{pbr_energy}, as demonstrated in Equation \ref{overall_energy_model}(c).
Once all the nodes transitioned into the Alert phase, the last node that has just transitioned may need up to $l \times 0.1024$ seconds to synchronize with all the nodes' wake-up time.
The energy consumed during this duration is demonstrated in Equation~\ref{overall_energy_model}(d).
Finally, $E_{smp}(t_{smp}) = t_{smp} \times \bar{E}_{smp}$, where $\bar{E}_{smp}$ is the energy consumed per second during the Sampling phase. 
We will present this energy in~\cref{energy_efficiency}.

\section{Packet Creation and Protocol Stack Processing}
\label{section:packet_processing}

Compared to the 802.15 family of standards, the protocol stack of 802.11 is more complicated and introduces unique challenges in terms of packet processing delay and its effect on the sampling rate and inter-sample interval.
In this section, we propose methods to tackle these challenges. 
We consider three IoT protocol stacks: LwIP~\cite{LwIP}, NetX~\cite{NetX}, and NetXDuo~\cite{NetXDuo}. 
LwIP is compatible with FreeRTOS, and NetX and NetXDuo work with ThreadX.
Although NetX and NetXDuo support UDP and TCP, the major difference is that NetXDuo supports both IPv4 and IPv6 stacks.
Considering the higher overhead of TCP, and since packets are transmitted over a single hop, we use UDP. 
We assume reliability is addressed by other means, such as link-layer retransmissions.

Despite using these operating systems and protocol stacks, the observations and proposed methods can be generalized to various systems, especially those including a single-core application processor.

\subsection{Packet Processing Delay and its Effect on Sampling Rate}
\label{pkt_proces_delay}
When samples and their timestamps are ready, there are three delay components incurred to transmit a packet:
packet preparation, network stack processing, and packet transmission by the transceiver.

Both FreeRTOS \cite{FreeRTOS,FreeRTOS_git} and ThreadX \cite{ThreadX,ThreadX_github} require the application thread to create a packet data structure.
This process, which we refer to as \textit{packet creation}, entails buffer allocation from a pool available by the network stack. 
Next, the application payload is sent to the network stack to compute and add proper headers, including TCP/UDP, IP, and MAC.
This process is referred to as \textit{packet processing}.
Finally, the frame is passed to the wireless subsystem for transmission.
Packet creation and packet processing tasks run on the application subsystem, and low-level packet transmission operations (e.g., carrier sensing, backoff) are handled by the wireless subsystem.
Therefore, it is essential to reduce the overhead of packet creation and processing to minimize their impact on the sampling task.

\begin{figure}[!t]
\centering
  \includegraphics[width=1\linewidth]{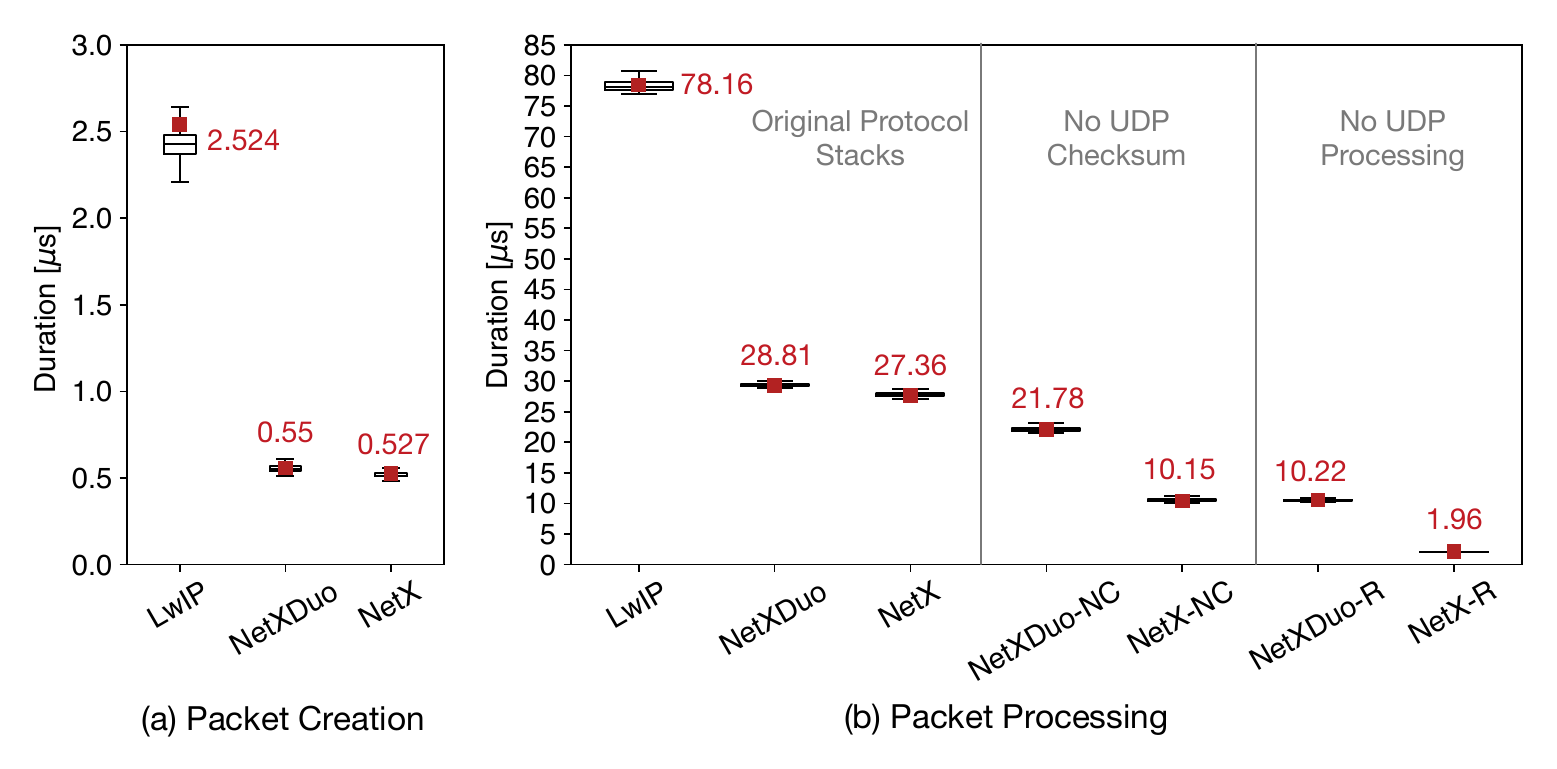}
  \vspace{-20pt}
  \caption{Empirical evaluation of (a) packet creation, and (b) packet processing duration. 
  The number on each box plot presents the mean value.
  NetX-NC and NetXDuo-NC refer to bypassing UDP checksum, where 'NC' stands for No Checksum.
  NetX-R and NetXDuo-R refer to \textit{raw} socket processing, which represents fully bypassing the UDP layer.}
  \label{fig:prtoc_stack_latency}
\end{figure}

Figure~\ref{fig:prtoc_stack_latency}(a) compares the packet preparation delay of LwIP, NetX, and NetXDuo.
The average {packet creation} delay is around $\mathrm{2.5\:\mu s}$ for LwIP and $\mathrm{0.5\:\mu s}$ for NetXDuo and NetX.
The main cause of this difference is the memory allocation method used for packet buffers.
With LwIP, the packet structure prepared in the application thread is copied to the network stack; whereas, NetX and NetXDuo provide a zero-copy method to prevent this overhead.

Once packet preparation is complete, the \texttt{packet\_send()} API is called (cf. Algorithm~\ref{alg:main_thread}).
To measure packet processing duration, we modified the network stack to keep track of two events for each packet: 
(\romannumeral 1) when the packet arrives in the network stack and 
(\romannumeral 2) when the packet is fully processed by the driver and is ready to be sent to the firmware (wireless subsystem).
Figure~\ref{fig:prtoc_stack_latency}(b) demonstrates packet processing delay.
In this figure, the name of a protocol stack refers to standard packet processing, including generating UDP, IP, and MAC headers.
The results demonstrate that the mean packet processing latency of NetX and NetXDuo is about $\mathrm{65\%}$ lower than that of LwIP.
One of the main reasons causing the poor performance of LwIP is that it performs multiple validation steps while the packet is passed through the stack.
For example, LwIP checks if the socket is valid before further packet processing.
Also, both NetX and NetXDuo implement an enhanced UDP processing implementation \cite{NetX,NetXDuo}.



To reduce protocol stack processing overhead, we modified NetX and NetXDuo to \textit{bypass UDP checksum generation}.
These enhanced stacks are denoted as NetX-NC and NetXDuo-NC, where 'NC' refers to No Checksum.
The results in Figure \ref{fig:prtoc_stack_latency}(b) show that bypassing UDP checksum reduces the packet processing time of NetX and NetXDuo by $\mathrm{62\%}$ and $\mathrm{24\%}$, respectively.

To further reduce packet processing delay, we fully bypass UDP processing. 
These protocol stacks are called NetX-R and NetXDuo-R, where 'R' stands for \textit{raw}.
With this enhancement, NetX-R and NetXDuo-R achieve $\mathrm{1.96\:\mu s}$ and $\mathrm{10.22\:\mu s}$ delay, respectively.
%
Considering the execution duration of these network stacks, when collecting 256-sample batches at $\textrm{500\:k\gls{sps}}$, the packet processing delay of LwIP, NetX, and NetXDuo result in missing about 39, 13, 14 samples per packet processing, respectively.
With NetX-R and NetXDuo-R, the number of missing samples is reduced to 1 and 5 samples per packet processing, respectively.
\textit{Considering the lower packet processing delay of NetX, we use this network stack in the rest of this paper.}



\subsection{Impact on Sampling Stability}
\label{pkt_proces_samplign_stability}
In \cref{encoding_methods} we assumed the only task being run by the processor is the sampling task, which is the task responsible for communicating with the ADC and transferring samples to the processor.
In this section, we use the term 'Sampling-Only' to refer to this scenario.
The results presented in \cref{encoding_methods} demonstrated that even the 'Sampling-Only' scenario results in inter-sample variations.
Since in the realistic scenarios the collected samples must be transmitted, in this section, we study how packet preparation, processing, and transmission exacerbate the variations of inter-sample intervals.
\begin{figure}[!t]
\centering
  \includegraphics[width=1\linewidth]{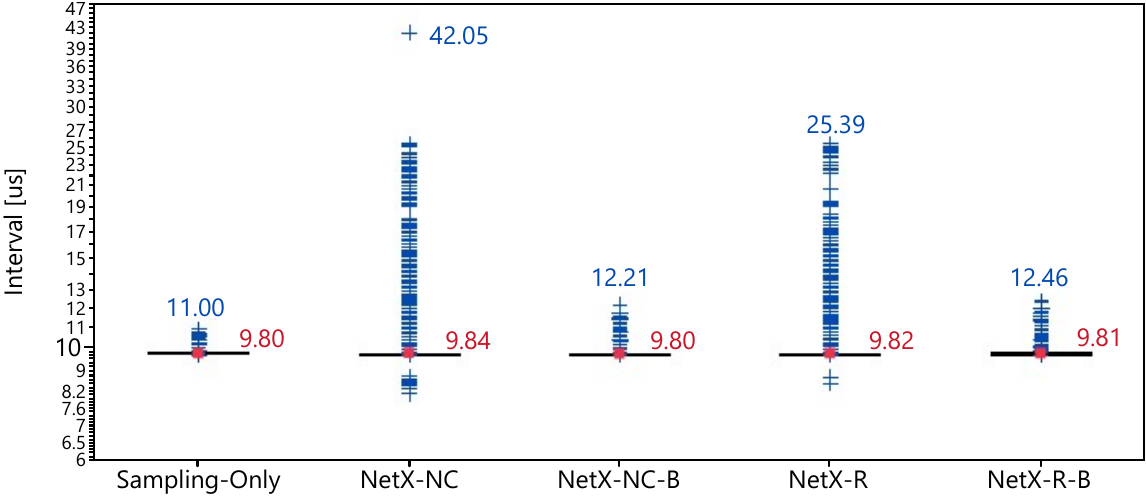}
  \caption{Variations of inter-sample intervals when performing 60000 iterations of collecting 512-sample batches.
  The sampling rate is $\textrm{100\:k\gls{sps}}$.
  The numbers on top of the outliers represent the maximum outlier value, and the numbers on the right side of each box plot represent the mean values.
  \textit{Sampling-Only} refers to the case where no packet processing task runs on the system.
  NetX-NC and NetX-R refer to \textit{no UDP checksum} and \textit{raw} packet processing, respectively.
  NetX-NC-B and NetX-R-B refer to the \textit{blocking call} versions of packet processing using NetX-NC and NetX-R, respectively.
  The blocking call implementations reduce interval variations by up to $\mathrm{70\%}$.
  Compared to Sampling-Only, the higher variations of NetX-NC-B and NetX-R-B are caused by the communication between driver and firmware.
  }
  \label{fig:semaphore}
\end{figure}
%
%

In a normal producer-consumer analogy, the sampling task (producer thread) passes the samples to the protocol stack (consumer thread) for processing.
Suppose the sampling and packet processing tasks are assigned the same priority level. 
In that case, the processor switches between the two tasks until the packet is fully processed, and both context switching and packet processing disturb the sampling task.
For example, in Figure \ref{fig:inter_sample_interval}(b), the processor switches to another task before collecting Sample 2.
In this case, the interval between the collection of Sample 1 and Sample 2 increases and results in higher differences between $t_2 - t_1$ and $t_3 - t_2$.

As Figure~\ref{fig:semaphore} demonstrates, the inter-sample intervals achieved by using NetX-NC and NetX-R are up to 42.05 and $\mathrm{25.39\:\mu s}$, respectively, which are significantly higher than Sampling-Only results.
These results confirm that running packet processing while the sampling task is in progress causes higher variations in inter-sample intervals.
An alternative is to assign a higher priority to the sampling task.
In this case, however, the high sampling rate and involvement of the processor in sample collection through the \gls{SPI} bus prevents the allocation of enough resources to packet processing, and this ultimately results in a buffer overflow.
The other alternative is to assign a higher priority to the packet processing task.
Although, in general, this configuration reduces the inter-sample variations, the packet processing task has the authority to interrupt the sampling task at any time. 
For example, when an incoming broadcast packet arrives, it causes context switching to the packet processing task, and the utilization of the processor by this task causes interruptions of the sampling task.

To remedy this problem, we modified the NetX stack by converting the \texttt{packet\_send()} API into a blocking call, while both sampling and packet processing tasks run at the same priority level.
The new API returns when the packet passes through all the layers and is processed by the driver.
With this method, outgoing packet processing runs sequentially after collecting each batch of samples, and packet processing receives full processor attention.
Also, in the case of incoming packets, the processor is shared between the two tasks.
These methods are named NetX-NC-B and NetX-R-B in Figure~\ref{fig:semaphore}, where `B' refers to \textit{blocking}.
As the results show, for NetX-NC-B and NetX-R-B, the maximum inter-sample variation is dropped by $\mathrm{70\%}$ and $\mathrm{50\%}$, respectively.
Note that converting \texttt{packet\_send()} to a blocking call does not mean that packets cannot be buffered.
We provisioned a 200-packet buffer in the driver.
Once the protocol stack fully processes a packet, the control returns to the sampling task if the buffer is not full.
If the buffer is full, the packet processing API waits for the packet buffer to become available before resuming sampling.

Even with using a blocking call, the range of variations is still higher than Sampling-Only.
We identified that the driver-firmware interactions cause these outliers.
The \gls{SoC} we used in our node design has two processors, as discussed in~\cref{sys_overview}. 
The packet exchange between the wireless and application processors is via \gls{DMA}.
Once the driver completes processing a packet, it is sent to the firmware immediately or later.
The delay of sending a packet to the firmware is because the firmware, which resides in the wireless subsystem, may be busy sending another packet.
Part of initiating a driver-firmware communication through \gls{DMA} requires attention from the application processor; therefore, the sampling rate is affected.
Nevertheless, these variations account for less than $\mathrm{1\%}$ of inter-sample intervals.

It is important to note that the inter-sample variations reported in this section are pertaining to the processor type used in our node design.
For example, a processor with a lower operating frequency may demonstrate higher variations due to the longer duration of packet processing.
Such studies across various hardware platforms are left as future work.

\section{Time Synchronization}
\label{time_sync}

Accurate analysis of the server's collected data requires establishing time synchronization across all the nodes and timestamping all the samples.
In the proposed system, we achieve time synchronization by the periodic transmission of a broadcast time synchronization packet from the server to all the nodes.
Each broadcast packet is sent once by the \gls{AP}, and all the nodes receive this packet concurrently.
When a node receives this packet, it modifies its internal timer by running a \gls{TSM} that adjusts the node's clock with the one received in the packet.
The time synchronization packet's payload size is nine bytes, where eight bytes represent nanosecond time and one byte is used for packet type identification.

Although, in general, establishing time synchronization in single-hop networks is more straightforward than in multi-hop networks, the use of 802.11 standard introduces unique challenges.
This section focuses on the delay of processing incoming packets as the leading cause of time synchronization inaccuracy.

\subsection{Packet Reception Delay}
\begin{figure}[!t]
\centering
  \includegraphics[width=1\linewidth]{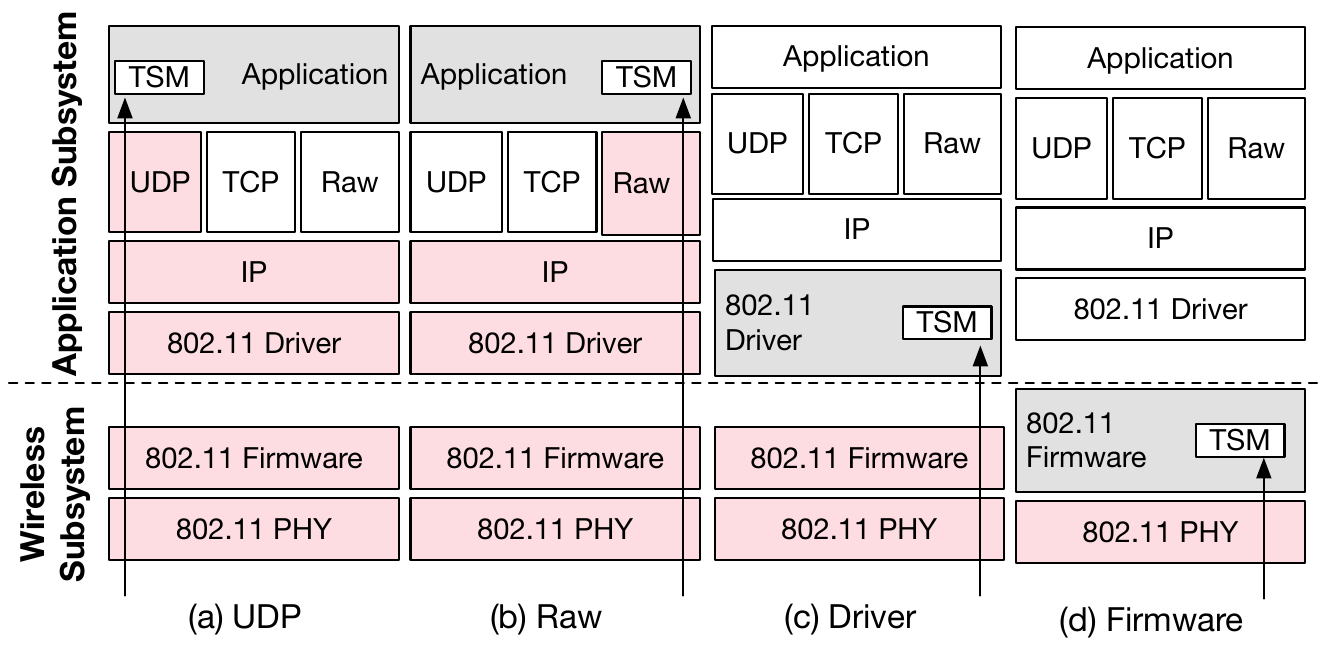}
  \vspace{-20pt}
  \caption{An incoming broadcast packet used for time synchronization passes through all the shaded layers until it reaches the layer that includes \gls{TSM}.
  }
  \label{fig:protoco_stack}
\end{figure}

\begin{figure}[!t]
\centering
  \includegraphics[width=1\linewidth]{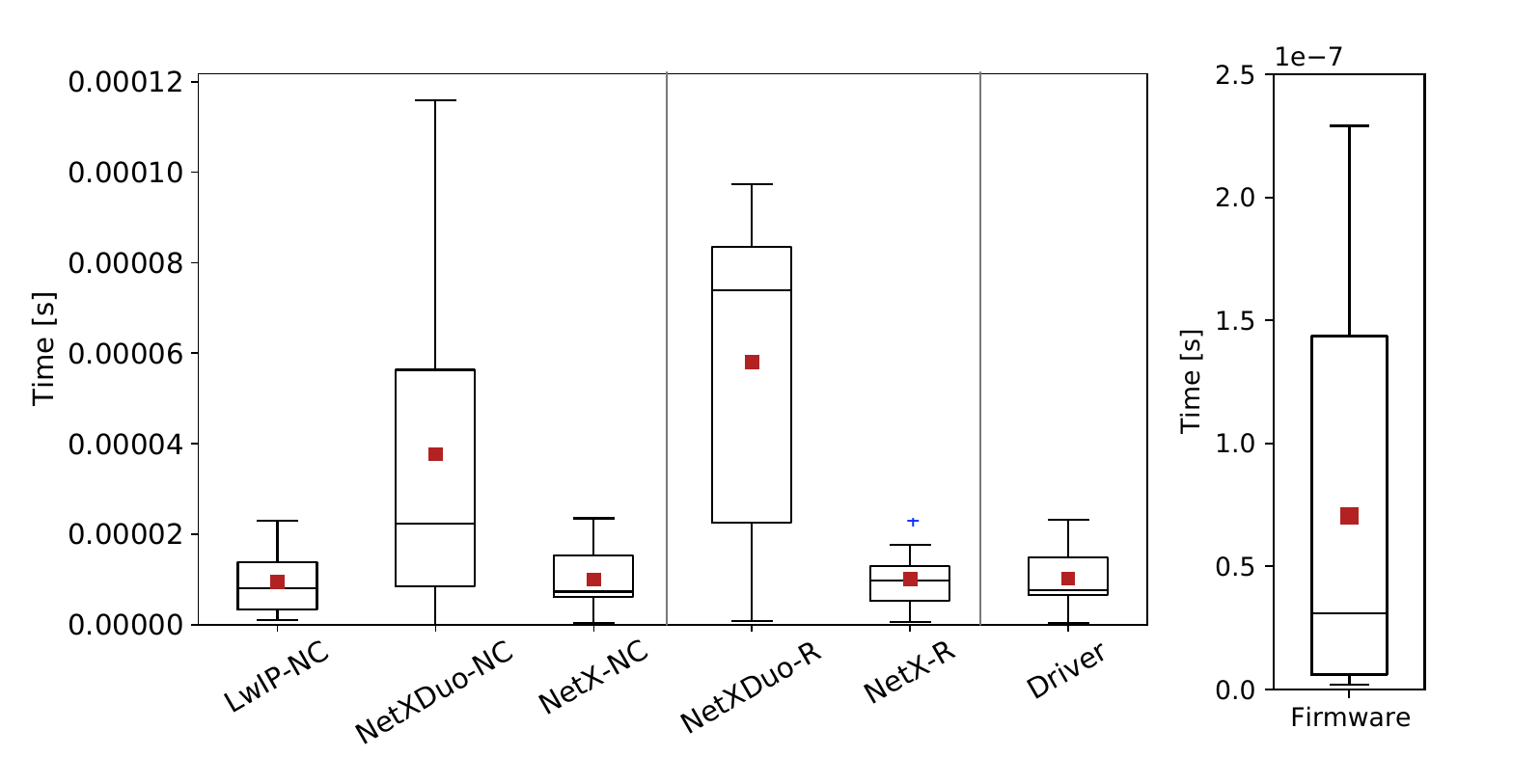}
  \vspace{-20pt}
  \caption{Empirical evaluation of time synchronization error.
  When the \gls{TSM} is implemented in the driver or a layer above it, the worst-case accuracy surpasses $\mathrm{20\:\mu s}$.
  When the \gls{TSM} is implemented in the firmware, the worst-case accuracy is less than $\mathrm{0.25\:\mu s}$.
  These results confirm that firmware-level time synchronization is required to achieve sub-$\mathrm{\mu s}$ accuracy.
  }
  \label{fig:time_sync_results}
\end{figure}

We study how protocol stack layers, from the 802.11 firmware up to the UDP layer, affect the delay of processing an incoming time synchronization packet.
We modified these network stacks and placed the \gls{TSM} in different layers.
Figure~\ref{fig:protoco_stack} shows these scenarios.
Since the ultimate goal of time synchronization is to ensure all the nodes use the same clock value when adding timestamps to samples, we measured the time difference between two nodes immediately after processing the time synchronization packet. 
This experiment was repeated for over one million iterations. 
Figure~\ref{fig:time_sync_results} presents the empirical results.
It is important to note that these results present the \textit{difference} between the delay of processing time synchronization packets by the two nodes in our testbed.
We detail these results as follows.
Bypassing UDP checksum (denoted as 'NC') in LwIP and NetX results in close to $\mathrm{30\:\mu s}$ error.
With NetXDuo, however, the max error is close to $\mathrm{120\:\mu s}$.
We identified this increase due to the higher processing overhead of NetXDuo caused by its IPv6 supporting feature.
The results also show that, although using raw sockets entirely bypasses UDP layer processing, it does not reduce time synchronization error.
The same observation is made even when the \gls{TSM} is placed in the driver.
These results indicate that the causes of processing time variability are located before the packet is delivered to the driver.
In particular, \textit{once subtracted across the two nodes}, the delay of memory allocation to the packet and processing it in the driver and upper layers is almost negligible compared to the delays caused before the packet arrives in the driver.
As discussed in~\cref{section:packet_processing}, packet exchange between the wireless and application subsystems is via \gls{DMA}.
With this architecture, time synchronization error is caused by the instance interrupts are raised by the wireless subsystem of each node.
Based on this observation, we moved the \gls{TSM} to the firmware, hence it is run by the wireless subsystem.
Thus, time synchronization happens before the received packet is transferred to the application subsystem via \gls{DMA}.
With this enhancement, as Figure~\ref{fig:time_sync_results} shows, the maximum error drops to less than $\mathrm{0.25\:\mu s}$.
This accuracy is sufficient to maintain time synchronization accuracy for sampling rates up to $\textrm{1\:Msps}$.

\subsection{Periodic Time Synchronization}
\label{per_time_sync}
The server must periodically transmit a time synchronization packet to ensure that the nodes' clocks do not deviate beyond a specific limit.
Determining the proper period depends on the clock stability of the nodes' processor.
The processor's clock frequency stability, denoted as $\nu$, varies based on changes in temperature, voltage, output load, and aging.
Frequency stability is typically expressed in parts per million (ppm). 
Peak frequency variation can be presented as $d_{f} = f \times \nu$, where $f$ is processor frequency~\cite{tirado2019performance}. 
The maximum time difference between two nodes per second can be expressed as $d_t = \frac{1}{(f-d_f)} - \frac{1}{(f+d_f)}$.
The CYW54907~\cite{CYW54907} used in our node design runs at $\textrm{160\:MHz}$ and its $\nu = \mathrm{\pm 2.5\:ppm}$. 
Therefore, its peak frequency variation is $d_{f} = \textrm{400\:Hz}$, and the maximum timing error per second between two nodes is $d_{t} = \mathrm{5\:\mu s}$.
%
%
The maximum synchronization error immediately after processing the time synchronization packet is less than $\mathrm{0.25\:\mu s}$. 
When sampling at $\textrm{500\:k\gls{sps}}$, a time synchronization error of less than $\mathrm{1\:\mu s}$ allows the server to correlate the samples, which are collected every $\mathrm{2\:\mu s}$.
To achieve a maximum error of $\mathrm{0.75\:\mu s}$, the period of time synchronization must be less than $\textrm{150\:ms}$.
For $\textrm{100\:k\gls{sps}}$, a time synchronization period of $\textrm{1.95\:s}$ limits the error to $\mathrm{9.75\:\mu s}$.

\section{Hardware Design} 
\label{hardware_design}
\begin{figure}[!t]
\centering
  \includegraphics[width=1\linewidth]{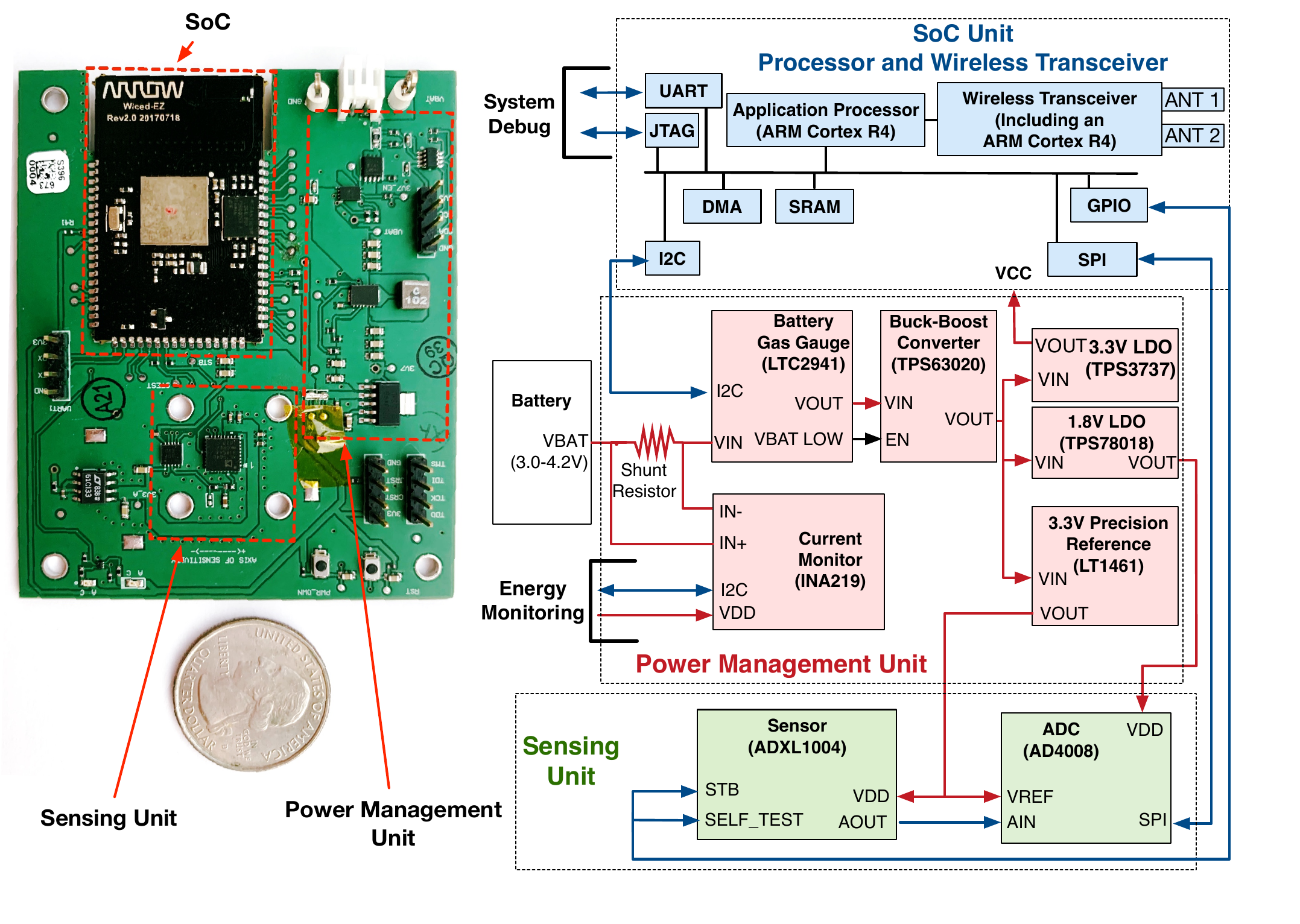}
  \caption{The (a) prototype and (b) block diagram of a \toolname{} node. 
  }
  \label{fig:wss_diagram}
\end{figure}

The primary hardware design considerations are low-power consumption and supporting high sampling rates.
%
%
Figure~\ref{fig:wss_diagram} presents the prototype and block diagram of a \toolname{} node.
At a high level, our proposed node design consists of three main units: 
(\romannumeral 1) The \gls{SoC} is used to collect data from the sensing unit, construct data packets, send packets to an \gls{AP}, and control the sensing and power management units.
(\romannumeral 2) The sensing unit is responsible for measuring acceleration forces and converting analog measurements to digital data.
(\romannumeral 3) The power management unit supplies all the components with a stable power source, provides an accurate reference voltage to the sensing unit, measures the node's power consumption, and protects the node from under-voltage discharge.
%
%
A node includes components (such as a current monitor, battery gas gauge, and probe pins) to facilitate performance study and adapt this design for various applications.

Since we have discussed \gls{SoC} in \cref{sys_overview}, we focus on the sensing and power management unit in this section.
Although this node is geared towards vibration test monitoring, it can be simply modified for various application types.

\subsection{Sensing Unit}
The sensing unit consists of two components, a \gls{MEMS} accelerometer and an \gls{ADC}.
The details of these components are discussed as follows. 

\subsubsection{Sensor}
Vibration test requires accelerometers to measure frequencies from $\mathrm{10\;Hz}$ to $\textrm{10\:kHz}$.
In addition to accuracy, the sensor must be low-power, low-cost, and available in a small form factor.
Considering these requirements, capacitive accelerometers are suitable for measuring static and dynamic accelerations, offering linearity, showing high output stability, and low cost.
The sensor used in our node design is ADXL1004~\cite{ADXL1004}, a \gls{MEMS} capacitive accelerometer with a measurement range of $\mathrm{\pm\:500}$ g-force and capable of measuring frequencies of up to $\textrm{24\:kHz}$. 
Its maximum power consumption is $\textrm{5\:mW}$ and its noise range is up to $\mathrm{125\:\mu g}/\sqrt{\mathrm{Hz}}$.
Sensor sensitivity is $\mathrm{2.64\:mV/g}$ with an input voltage of $\textrm{3.3\:V}$.
To achieve this sensitivity, we use a precision voltage rail for the sensor's power input to provide a low variance ($\mathrm{\leq\:1\%}$) across temperature and node-to-node.


\subsubsection{ADC}
\label{hardware_adc_section}
Since the highest sampling frequency measured by the sensor is $\textrm{24\:kHz}$, according to the Nyquist theorem, the sampling rate must be at least $\mathrm{2\times 24\:\mathrm{kHz}}$.
The sensor has $\mathrm{2.64\:mV/g}$ sensitivity with input voltage $\textrm{3.3\:V}$.
Although a 12-bit \gls{ADC} with $\textrm{1.6\:mV}$ resolution and input voltage of $\textrm{3.3\:V}$ is sufficient to acquire the sensor's output accurately, we use an \gls{ADC} that provides higher resolution and is capable of sampling at considerably higher rates.
With this design, the \gls{ADC} can interface with different types of sensors and be used in applications with higher demands.
Specifically, our design includes a $\textrm{16-bit}$ AD4008~\cite{AD4008} that supports a maximum sampling rate of $\textrm{500\:k\gls{sps}}$. 
To interface the \gls{ADC} with the \gls{SoC}, we developed a custom \gls{SPI} bus driver.
To this end, in addition to the datasheet, we used the FPGA-based evaluation board and the Windows-based software available for this \gls{ADC} to understand its timing requirements. 

The CONV pin of AD4008 is not an interrupt-driven method to detect conversion completion; instead, this pin is used to trigger conversion.
Although some \glspl{ADC} (e.g., ADS8166 and MAX12555) provide a pin to detect sample conversion time, the processor involvement is necessary to send and receive messages over the SPI bus (as discussed in \cref{pkt_proces_samplign_stability}).
Although some \glspl{SoC} include \gls{ADC} and support DMA transmission to the main memory, they do not support 802.11 communication, and therefore it is necessary to transfer the data to another SoC; however, this process introduces additional overhead.
We leave the design, development, and potential benefits of such node architecture as future work.

To prevent aliasing and non-linear charge kickback at the start of each ADC acquisition phase, we placed a first order anti-aliasing RC filter at the ADC's input. 
In addition, AD4008 provides a high-z mode feature that mitigates the effects of non-linear charge kickback and reduces distortion over a wide frequency range up to $\textrm{100\:kHz}$.


\subsection{Power Management Unit}
The power management unit must be capable of operating over a wide load profile. 
%
The gas gauge is used to protect the battery (lithium-polymer) from under-voltage discharge and monitor the battery's remaining charge. 
%
The bucket-boost converter (TPS63020) is part of the node's main voltage rail and feeds into the cascaded linear regulators. 
By simulating performance across the design input voltage range and load currents up to $\textrm{500\:mA}$, the bucket-boost converter shows an efficiency range between $\mathrm{88\%}$ at light current load and up to $\mathrm{94.5\%}$ for the max current load~\cite{TPS6302x}. 
However, the bucket-boost converter introduces a switching noise that affects other sensitive analog components when the performance efficiency reaches $\textrm{90\%}$.
To reduce the noise and provide a steady voltage output, our design uses TPS3737~\cite{TPS737xx} $\textrm{3.3\:V}$ and TPS78018~\cite{TPS780xx} $\textrm{1.8\:V}$ \gls{LDO} regulators.
The $\textrm{3.3\:V}$ regulator provides a stable voltage for the processor and wireless transceiver unit, and the $\textrm{1.8\:V}$ regulator provides a stable voltage for the \gls{ADC}.
%
%
%
%
Since AD4008 and ADXL1004 only consume a maximum of $\textrm{2\:mA}$ during sampling, the impact of dissipated power loss for the $\textrm{1.8\:V}$~\gls{LDO} regulator is negligible.

Our design uses a LT1461~\cite{LT1461} to provide an extremely constant voltage reference that is resistant to temperature, noise on the supply rail, and process variations~\cite{voltagereference}. %
We connect the LT1461 to the ADXL1004 and AD4008 to prevent the risk of switching noise or voltage ripple, thereby reducing measurement inaccuracies.
It can also deliver up to $\textrm{50\:mA}$, and the output voltage is guaranteed to be within $\mathrm{\pm0.08\%}$ of the nominal value with a temperature drift of only $\mathrm{12\:ppm/^{\circ}C}$.
For developmental purposes and to eliminate the need to use an external power meter, an INA219~\cite{dezfouli2018empiot} is used for in-circuit power measurement.

\label{sec:power_management_unit_energy}
We used a high-precision \gls{DMM} \cite{k7510} to study the power consumed by the components of the power measurement unit.
Table~\ref{table:pahse_active_comp} presents the results.
\begin{table}[]
\centering
\setlength{\tabcolsep}{0.6em} 
\caption{Current consumption by the power management unit}
\begin{tabular}{c|c|c|c|c|c|c|}
\cline{2-7}
                                                                                    & \multicolumn{5}{c|}{Active Component's Current (mA)}                                                                                                                                                                                                                                                          & \multirow{2}{*}{\begin{tabular}[c]{@{}c@{}}Total\\ Current\\(mA)\end{tabular}} \\ \cline{2-6}
                                                                                    & \begin{tabular}[c]{@{}c@{}}Battery \\ Gas\\ Gauge\end{tabular} & \begin{tabular}[c]{@{}c@{}}Buck\\ Boost\\ Converter\end{tabular} & \begin{tabular}[c]{@{}c@{}}3.3V \\ LDO\end{tabular} & \begin{tabular}[c]{@{}c@{}}1.8V \\ LDO\end{tabular} & \begin{tabular}[c]{@{}c@{}}3.3V\\ Ref \end{tabular} &                                                                          \\ \hline
\multicolumn{1}{|c|}{\begin{tabular}[c]{@{}c@{}}Idle\textbackslash Alert \end{tabular}} & 0.1  & 0.105  & 0.297  &   -  &  -    &  0.502      \\ \hline
\multicolumn{1}{|c|}{\begin{tabular}[c]{@{}c@{}}Sampling \end{tabular}}      & 0.1  & 0.105  & 0.297  &   0.872   &  0.07 &  1.444       \\ \hline
\end{tabular}
\label{table:pahse_active_comp}
\end{table}
During the Idle phase and Alert phase, both the ADXL1004 and AD4008 are completely turned off, and only the \gls{SoC} needs to be powered.
Table~\ref{table:pahse_active_comp} shows that the current consumption of the power management unit during these phases is $\mathrm{502\:\mu A}$.
%
The sensor and \gls{ADC} must be turned on during the Sampling phase.
Due to the activation of the $\textrm{1.8\:V}$ \gls{LDO} and the $\textrm{3.3\:V}$ reference voltage, the power consumption of the power management unit increases to $\textrm{1.444\:mA}$.
We present the overall energy consumption of a node during the Sampling phase in \cref{energy_efficiency}.

\section{Overall System Evaluation}
\label{perf_eval}
In this section, we evaluate the overall performance of the proposed system in terms of sampling rate, data generation rate, and energy efficiency. 

\subsection{Effective Sampling Rate and Data Rate}
\label{eff_samp_rate}

\begin{figure}[!t]
\centering 
\subfloat[\scriptsize 256-Sample Batch\label{pro_a}]{
\includegraphics[width=0.475\linewidth]{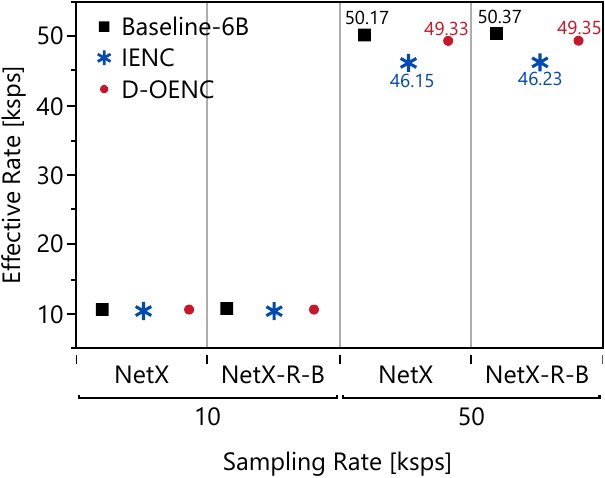}

}
\subfloat[\scriptsize 256-Sample Batch\label{pro_b}]{
\includegraphics[width=0.475\linewidth]{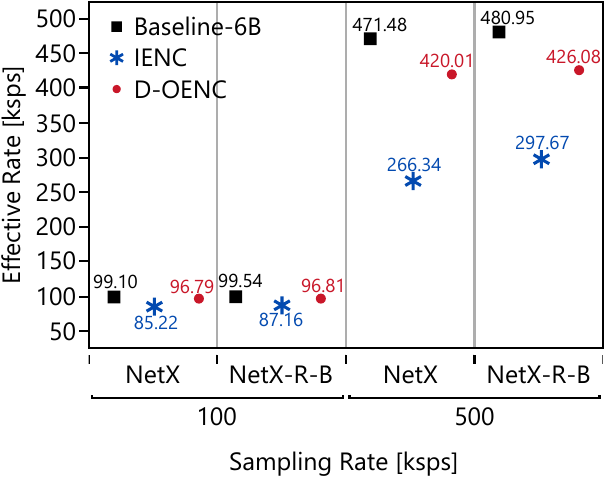}

}
\\
\subfloat[\scriptsize 512-Sample Batch \label{eva_a}]{
\includegraphics[width=0.475\linewidth]{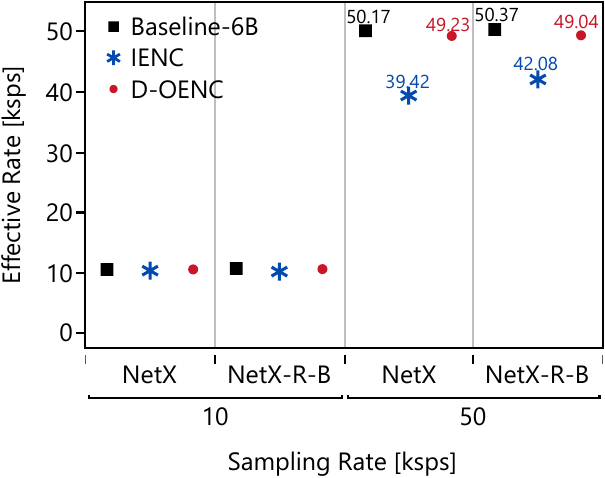}

}
\subfloat[\scriptsize 512-Sample Batch\label{eva_b}]{
\includegraphics[width=0.475\linewidth]{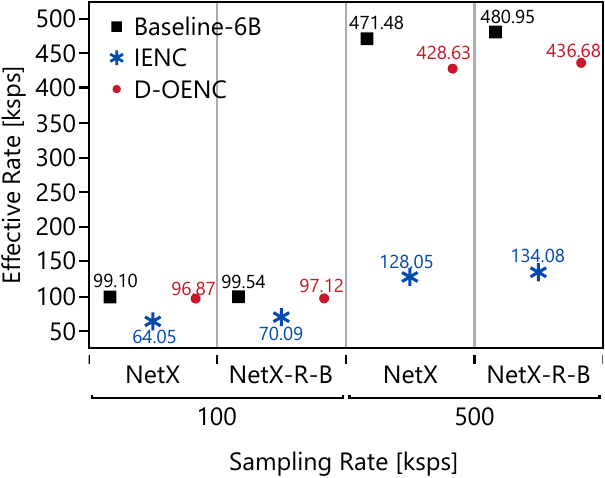}
\hfill
} 
\caption{Empirical measurement of \textit{effective rate} for (a) and (b): 256-sample batches, and (c) and (d): 512-sample batches.
The effective rate is the actual number of samples collected per second by a node.
Note that the number of samples per packet when using Baseline-6B is always 183, independent of batch size.
NetX refers to the default protocol stack, and NetX-R-B refers to the blocking call implementation of raw packet processing.
}
\label{fig:duration}
\end{figure}

\begin{figure}[!t]
\centering 
\subfloat[\scriptsize 256-Sample Batch\label{1da}]{
\includegraphics[width=0.475\linewidth]{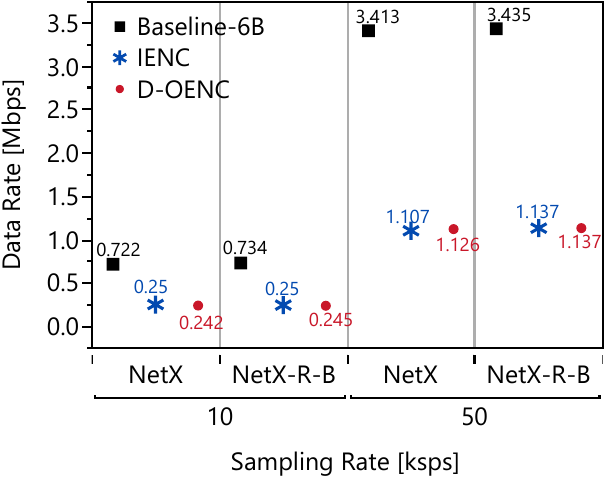}
\hfill
}
\subfloat[\scriptsize 256-Sample Batch\label{2da}]{
\includegraphics[width=0.475\linewidth]{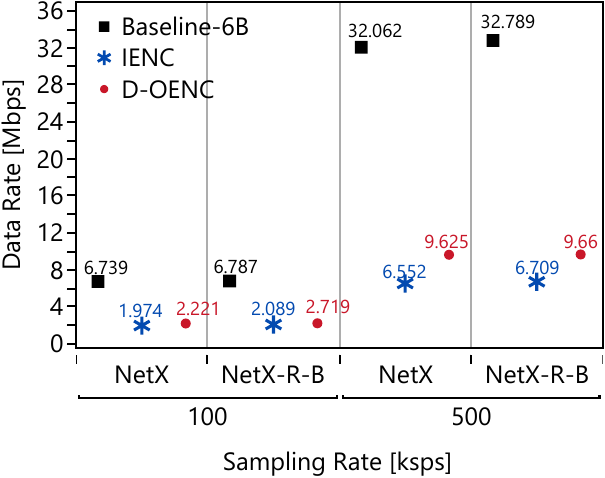}
\hfill
}

\subfloat[\scriptsize 512-Sample Batch\label{3da}]{
\includegraphics[width=0.475\linewidth]{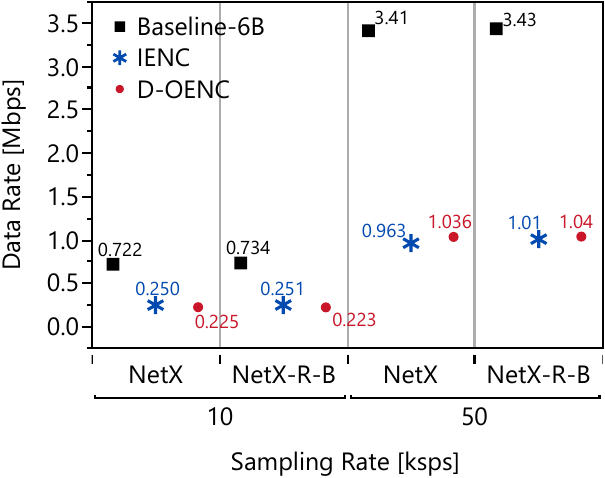}
\hfill
}
\subfloat[\scriptsize 512-Sample Batch\label{4da}]{
\includegraphics[width=0.475\linewidth]{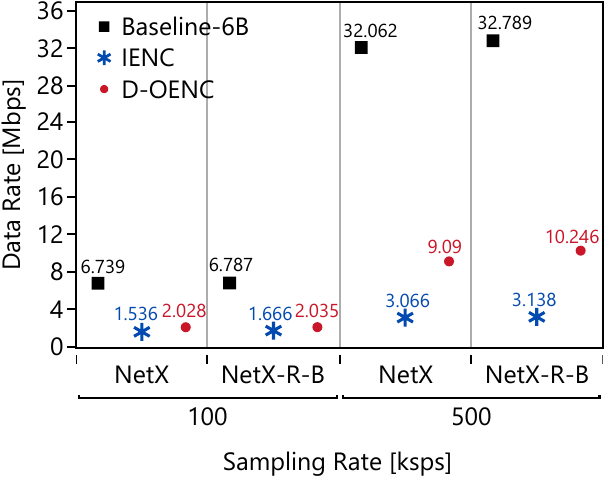}
\hfill
} 
\caption{Empirical measurement of wireless communication rate.
The data generated per second by a node includes samples, timestamps, packet encoding data, and packet headers.
Note that the number of samples per packet when using Baseline-6B is always 183, independent of batch size.
}
\label{fig:data_rate}
\end{figure}

During the Sampling phase, each node switches between sampling, packet encoding, packet creation, and packet processing.
To quantify the processing overhead of these tasks on the sampling rate, we define \textit{effective rate} as the actual number of samples collected and transmitted per second by a node.
Figure~\ref{fig:duration} shows the empirical evaluation of the {effective rate}.
On the other hand, to quantify the effectiveness of the encoding methods, Figure \ref{fig:data_rate} presents the empirical measurement of wireless communication rate.
%
%

Both Figures~\ref{fig:duration} and \ref{fig:data_rate} include Baseline-6B, which represents the case where no encoding is applied on timestamps (\cref{encoding_methods}).
Note that the number of samples per packet when using Baseline-6B is always 183, independent of batch size.
As Figure~\ref{fig:duration} shows, we observe that the sampling rate achieved by Baseline-6B is closer to the target rate, compared to the cases where encoding is applied.
This is because packet encoding introduces a processing time that causes sampling interruption.
However, as Figure \ref{fig:data_rate} shows, the data rate generated per node is considerably higher when using Baseline-6B.
For example, when sampling at $\textrm{500\:k\gls{sps}}$, the sampling rate of \gls{D-OENC} is $\mathrm{10\%}$ lower than that of Baseline-6B (Figure \ref{fig:duration}(d)), whereas, the data rate generated by Baseline-6B is $\mathrm{220\%}$ higher (Figure \ref{fig:data_rate}(d)).
Regardless of the channel access mechanism, the significantly higher data rate of Baseline-6B increases the number of \glspl{AP} required to ensure the bandwidth and reliability requirements of communication are met.
Also, note that for sampling rates less than $\textrm{500\:k\gls{sps}}$, the effective rate achieved by \gls{D-OENC} is almost the same as the target rate, but the amount of data generated by Baseline-6B is considerably higher.

As the sampling rate increases, the duration required to collect a batch drops, but the overhead of encoding and packet processing remains unchanged; thereby, the effective rate drops as the sampling rate increases.
With \gls{D-OENC}, although the results in~\cref{encoding_exec_time} demonstrated a $\mathrm{147\%}$ increase in processing time when switching from 256- to 512-sample batch, the results show almost similar performance for the two batch sizes.
There are two reasons for this observation:
First, packet processing overhead halves when the batch size doubles.
Second, due to the \gls{SPI} bus initialization time, the time it takes to collect 512 samples consecutively is less than collecting this number of samples in two 256-sample batches.
In summary, by using \gls{D-OENC} and NetX-R-B, we are able to collect up to $\textrm{436\:k\gls{sps}}$ when the \gls{ADC} operates at $\textrm{500\:k\gls{sps}}$, as Figure~\ref{fig:duration}(d) shows.
With \gls{IENC}, however, we observe a considerably lower effective rate when comparing the smaller and larger batch sizes, i.e., Figures~\ref{fig:duration}(a) and (b) against (c) and (d).
Based on the results presented in \cref{encoding_exec_time}, increasing the batch size from 256 to 512 exhibits a $\mathrm{376\%}$ increase in the execution time of \gls{IENC}.
This significant increase in encoding time dominates the overheads caused by packet processing and \gls{SPI} bus initialization.

These results also show the effect of network stack on both effective rate and data rate.
Compared to NetX, using NetX-R-B results in faster packet processing, thereby increasing the effective rate.
On the other hand, since NetX-R-B bypasses UDP processing, the packets do not include the UDP header, hence reducing the amount of data generated per second.
For example, when sampling at $\textrm{500\:k\gls{sps}}$, Figure~\ref{fig:duration}(b) shows that using \gls{D-OENC} with NetX-R-B increases the number of samples collected per second by about 6000, compared to NetX.
This comes at $\textrm{0.025\:Mbps}$ increase in data rate, as Figure~\ref{fig:data_rate}(b) demonstrates.
With \gls{IENC}, Figure~\ref{fig:data_rate}(b) shows that sampling rate increases by about 31000, and this causes $\textrm{0.157\:Mbps}$ increase in data rate.

\subsection{Energy Consumption During Sampling Phase}
\label{energy_efficiency}
\begin{figure}[!t]
\centering 
\subfloat[\scriptsize 256-Sample Batch\label{cb_a}]{
\includegraphics[width=0.475\linewidth]{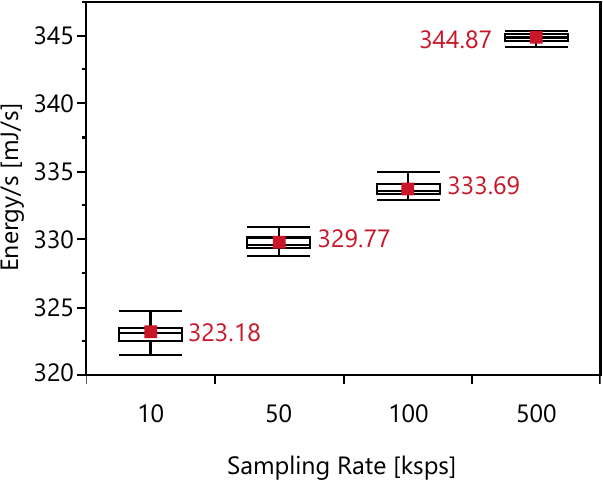}
\hfill
}
\subfloat[\scriptsize 512-Sample Batch\label{cb_b}]{
\includegraphics[width=0.475\linewidth]{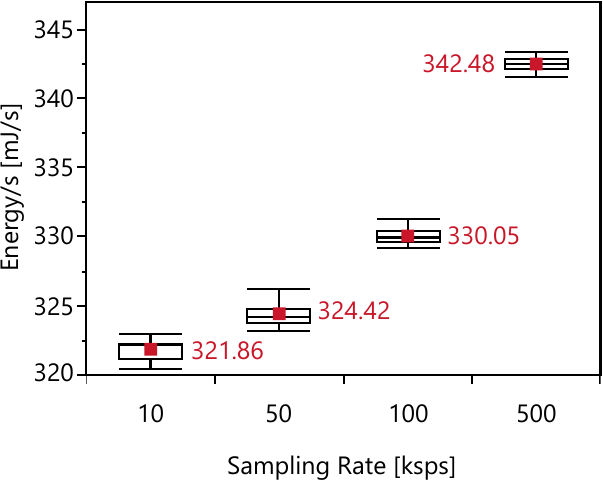}
\hfill
} 
\caption{Empirical measurement of energy consumption during the Sampling phase.
While the effective rate of the 512-sample batch size is higher (Figure \ref{fig:duration}), its energy consumption is lower because of the lesser overhead of packet processing and transmission.
}
\label{fig:power_measurement}
\end{figure}

This section presents the energy consumption of a node that uses \gls{D-OENC} and NetX-R-B during the {Sampling phase}.
Figure~\ref{fig:power_measurement} shows the empirical measurement of energy consumption for various sampling rates.
As we increase the sampling rate from $\textrm{10\:k\gls{sps}}$ to $\textrm{500\:k\gls{sps}}$, the energy consumed per second increases by $\mathrm{6.7\%}$ and $\mathrm{6.4\%}$ when using 256- and 512-sample batches, respectively.
The higher energy consumption when using 256-sample batches is due to the higher packet header overhead caused by sending more packets.
Also, with the smaller batch size, packet processing overhead increases and the number of switching between reception and transmission mode doubles.

\section{Related Work}
\label{relatework}

The existing \gls{SHM} systems are built on top of the $\textrm{802.15}$ standards to reduce installation and maintenance costs for structural monitoring~\cite{whelan2009real,samuels2011wireless,lynch2004design}, machine health monitoring~\cite{bengherbia2017fpga,huang2015development}, space applications~\cite{swathy2014analysis,dodson2018high}, and natural hazard monitoring~\cite{girard2013environmental,weber2012design}.
These standards offer low energy consumption, low cost, ease of integration with various microprocessors, and the simplicity of configuring and customizing physical layer and MAC parameters~\cite{dodson2018high,phanish2015wireless,whelan2009real,dezfouli2015modeling,dezfouli2014cama}.
However, these systems do not support ultra-high-rate applications; instead, they either focus on the networking and energy efficiency aspects of multi-hop communication or propose methods for distributed processing and data aggregation~\cite{alves2017damage}.
%
%
Although using distributed processing can reduce communication rate demand, not all applications can benefit from this method.
The applications that require long sampling duration or raw sample collection across nodes (similar to that discussed in \cref{app_scenario}) require centralized processing to enhance analysis accuracy and reduce the processing burden of nodes ~\cite{mechitov2004high,kim2007health}.

\label{subesection:802.11}
Abedi et al.~\cite{abedi2019wi} showed that at the physical layer, the energy consumption of 802.11 is $\textrm{10\:-\:100\:nJ/bit}$, while the range is $\textrm{275\:-\:300\:nJ/bit}$ for \gls{BLE}.
They also argued that the overall energy consumption of 802.11 is higher than \gls{BLE} due to the overheads pertaining to association and maintaining association.
The benefits of using 802.11 for building high-rate, energy-efficient \glspl{WSN} have been studied recently.
Tramarin et al.~\cite{tramarin2015use} evaluated the suitability of 802.11n in industrial applications from its physical layer point of view.
%
Luvisotto et al. \cite{luvisotto2016ultra} provided an overview of the requirements of industrial wireless networks and studied the suitability of 802.11 to address these requirements, compared to cellular and 802.15 networks.

Yu et al.~\cite{yu2015study} proposed an 802.11-based \gls{WSN} to monitor potential damages to bridge parts while they are being lifted and mounted.
The sampling rate of each node is $\textrm{100\:\gls{sps}}$, each packet contains $\textrm{80\:samples}$, and the size of each sample is $\textrm{16\:bits}$.
%
%
RT-WiFi \cite{wei2013rt,tramarin2019real} can sample at up to $\textrm{6\:ksps}$.
The basic idea is to modify the driver and mac80211 module on a Linux machine and enforce transmission schedules received from a central controller. 
They also evaluated system performance in a gait rehabilitation application where a mobile assistive robot and a smart shoe collect and transmit samples at $\textrm{100\:sps}$ and $\textrm{1\:ksps}$, respectively, and the controller replies at $\textrm{100\:sps}$.
Their work, however, does not address the challenges pertaining to scalability, high-rate transmission, and energy efficiency.
%
Dombrowski and Gross \cite{dombrowski2015echoring} proposed a distributed and deterministic channel access method to enhance the reliability of using token-ring for channel access arbitration.
They utilize software-defined radios operating in a $\textrm{10\:MHz}$ channel. 
The sampling rate of this solution is $\textrm{20\:sps}$.
%
Khorov et al. \cite{khorov2020modeling} presented the periodic reservation methods available in various 802.11 standards including 802.11s, 802.11ad, 802.11ax, and 802.11be.
They also addressed the challenges of establishing communication reservation periods for exchanging real-time multimedia traffic between two stations in the presence of interference.
The use of 802.11 standards in real-time multimedia applications has been studied in \cite{blanes2015802} and \cite{santonja2010analysis}.
Seno et al. \cite{seno2016enhancing} proposed a centrally-controlled 802.11 network to decide the admission of nodes that require real-time communication.
They use an \gls{EDF} scheduling method to admit nodes into the network, subject to their deadline and retransmission requirements.
Their proposed method has been implemented on the Linux operating system and the ath9k driver.
They present a thorough analysis of reliability; however, there is no study of the sampling rate.
Bartolomeu et al. \cite{bartolomeu2018supporting} addressed the challenges of transmitting real-time packets in the presence of non-real-time traffic.
%

In summary, the sampling rates achieved by the aforementioned works are far below the requirements of ultra-high-rate applications.
Also, in contrast to our work, the use of extended-period beacon reception and periodic association has not been studied in the existing works.
In particular, existing studies are primarily focused on reducing communication overhead in smartphones, and none of them propose methods for applications with long inactivity periods.


In this paper, we underlined the importance of time synchronization for accurate sample timestamping.
Achieving accurate time synchronization across nodes is also essential for TDMA-based channel access and real-time communication, and therefore, it has been studied by existing works \cite{wei2013rt,uchimura2010ieee,mahmood2016clock,tramarin2019real,sheu2007clock,zhou2005compatible}.
$\textrm{IEEE\:802.11}$ specifies a \gls{TSF} to time synchronize the wake-up instances of stations for beacon reception.
However, as per industrial implementations, the accuracy of this method is within $\mathrm{25\:\mu s}$; thereby, it cannot be used in high-rate applications.
%
%
Sevani et al. \cite{sevani2012implementation} utilized Madwifi driver (for Linux) and employed hardware packet timestamping to achieve a synchronization error of $\mathrm{10\:\mu s}$.
Uchimura et al.~\cite{uchimura2010ieee} considered using \gls{TSF} for time synchronization in adhoc 802.11 networks.
They analyzed the effect of channel access contention and beacon collision on accuracy.
Their empirical results using RT-Linux show the clock accuracy of $\mathrm{27\:\mu s}$. 
Wei et al. \cite{wei2013rt} modified the ath9k driver on Linux to utilize \gls{TSF} for achieving time synchronization in a single-hop network.
Assuming beacon transmission every 100 ms, they consider a maximum clock drift of $\mathrm{20\:\mu s}$.
To achieve nanosecond precision synchronization, Seijo et al. \cite{seijo2020enhanced} proposed a timestamping method that, in particular, takes into account the packet arrival rate caused by the multipath effect.
They utilize simulation-based studies to evaluate the proposed method.
In short, the existing time synchronization methods (e.g., $\mathrm{20\:\mu s}$ in \cite{zhou2005compatible} and \cite{wei2013rt}, $\mathrm{27\:\mu s}$ in \cite{uchimura2010ieee}, and $\mathrm{10\:\mu s}$ in \cite{sevani2012implementation}) do not provide the accuracy required for sample timestamping in ultra-high-rate applications.

Although modification of the 802.11 stack and bypassing layer-3 and layer-4 processing are the most common approaches employed to develop 802.11-based \glspl{WSN} and enhance time synchronization accuracy, there is no study concerning the effect of packet processing on sampling jitter.
Furthermore, in contrast with the existing work that utilize the Linux operating system (e.g., \cite{wei2013rt,uchimura2010ieee,seno2016enhancing,mahmood2016clock,tramarin2019real,mahmood2014delay}), in this paper we utilized \glspl{RTOS} and lightweight protocol stacks primarily designed for IoT applications.



Some WSN designs employ non-live data collection to cope with the low transmission rate of 802.15.4 links or reduce the effect of wireless transmission on the sampling rate.
For example, in \cite{kim2007health}, each node stores the collected samples in its local storage, and then the saved data is sent over $\textrm{802.15.4}$ links.
The impact of writing to flash storage on the stability of sampling rate was studied in \cite{kim2007health}. 
They demonstrated that when sampling at $\textrm{5\:k\gls{sps}}$, the sampling jitter is increased by $\mathrm{10\:\mu s}$.
They argued that a multi-processor node design is required to tackle this challenge.
Dai et al. \cite{dai2017wireless} showed that the process of sample collection and writing to a memory card reduces the sampling rate.
To address this problem, they used the \gls{DMA} controller to directly transfer data from \gls{ADC} to SRAM.
The user can then request for wireless transfer of the collected samples.
Although they show the enhancements in terms of the processor cycles saved, no evaluation of the actual sampling rate was presented.
Phanish et al. \cite{phanish2015wireless} also used \gls{DMA} to enhance sampling rate stability.
Specifically, each sampling round collects and transfers $\textrm{16}$ samples to SRAM via \gls{DMA}. 
Then, each batch of samples is sent as one packet.
Compared to our work, none of the existing works studied the effect of packet preparation and transmission on sampling rate.



In addition to non-live data collection, various compression methods have been used to reduce the amount of data transmitted per node.
One common approach for time-series-based data is to compute auto-regressive/auto-regressive moving average (AR/ARMA) coefficients locally by each node and transmit this data instead of the raw sensor data~\cite{chintalapudi2006monitoring,liu2011energy}, or use AR methods with random decrements to compress sensor data by averaging a large number of time segments~\cite{hu2012cluster}.
Nevertheless, these approaches are lossy and cannot be used in scenarios where raw data collection is required to implement various data analysis methods~\cite{liu2016senetshm,werner2006deploying}.
Furthermore, even if a drop in monitoring accuracy is tolerated, the compression level provided by these methods is not enough to adapt low-rate wireless technologies for ultra-high-rate applications.
Finally, the high processing complexity of these compression algorithms directly affects the sampling rate.
Another approach is to process data locally and send samples only if a particular event is detected~\cite{ling2009localized,werner2006deploying}.

\section{Conclusion}
\label{conclusion}
This paper presented \toolname{}, a system based on the 802.11 (WiFi) standard for ultra-high-rate sensing applications.
Our main observations and contributions are as follows:
(i) We showed that timestamp encoding is necessary to reduce the overhead of data transmission. 
We proposed a lossless encoding method that achieves a high compression rate and low processing overhead.
(ii) Considering the unique operational phases of the 802.11 standard, we proposed methods to enhance network longevity.
In particular, we showed that periodic reassociation with the \gls{AP} may be preferred over long beacon reception intervals.
(iii) We profiled the overhead of various network stacks in terms of outgoing packet preparation, processing, and their impact on sampling rate and stability.
We significantly reduced packet processing overhead by bypassing the UDP layer.
Also, by implementing a blocking packet processing call, we minimized the effect of packet processing on sampling stability.
(iv) We studied the delay of processing time synchronization packets from firmware to the application layer, and we proved that firmware-based processing is necessary to achieve a sub-$\mathrm{\mu}$s accuracy.
(v) We proposed a low-power node design for ultra-high-rate applications and presented empirical micro-benchmarks and overall system evaluation results using this node design.
Our work leveraged the two widely used \glspl{RTOS} (FreeRTOS and ThreadX) and three low-power network stacks (NetX, NetXDuo, and LwIP).


Although \toolname{} provides live data transmission, no per-packet delivery delay is enforced.
To enable the adoption of the proposed work in applications where real-time data delivery is a requirement~\cite{dezfouli2017rewimo}, an area of future work is to establish a time-slotted channel access scheduling among nodes.
In this work, we enhanced scalability by reducing the amount of data generated per node.
An area of future work is to evaluate system scalability and communication reliability using a large testbed and considering various channel access methods.
Compared to 802.11ac used in this paper, the 802.11ax standard offers OFDMA, which can be leveraged for the allocation of sub-carriers to nodes based on their buffered data.
In addition, MU-MIMO allows the AP to concurrently receive and decode signals received from multiple nodes.
Both of these methods enhance the packet delivery rate and reduce the energy consumption of nodes.
Such evaluations require a large-scale testbed and alternative hardware platforms. 
Novel methods must also be developed for systems that include more than one \gls{AP}.
For example, managing the association point of each node can be leveraged to reduce the number of retransmissions and enhance system reliability.

\ifCLASSOPTIONcaptionsoff
  \newpage
\fi



%


\bibliographystyle{IEEEtran}
\bibliography{bibliography}

%








\end{document}